\documentclass[12pt]{article}
\usepackage{amsmath}
\usepackage{graphicx}
\usepackage{hyperref}
\usepackage[]{natbib}
\usepackage{url} 
\newcommand{\blind}{0}

\usepackage{caption}
\usepackage{subcaption}

\usepackage{dsfont}
\usepackage{amsthm}
\usepackage{amsfonts}
\usepackage{mathtools}

\usepackage{caption}
\usepackage{subcaption}

\newtheorem{definition}{Definition}
\newtheorem{lemma}[definition]{Lemma}
\newtheorem{theorem}{Theorem}

\usepackage{algorithm}
\usepackage{algpseudocode}

\algnewcommand{\Inputs}[1]{%
  \State \textbf{Inputs:}
  \Statex \hspace*{\algorithmicindent}\parbox[t]{.8\linewidth}{\raggedright #1}
}
\algnewcommand{\Initialize}[1]{%
  \State \textbf{Initialize:}
  \Statex \hspace*{\algorithmicindent}\parbox[t]{.8\linewidth}{\raggedright #1}
}
\algnewcommand{\Returns}[1]{%
  \State \textbf{Returns:}
  \Statex \hspace*{\algorithmicindent}\parbox[t]{.8\linewidth}{\raggedright #1}
}

\usepackage{tikz}
\usetikzlibrary{shapes,decorations,arrows,calc,arrows.meta,fit,positioning}
\tikzset{
    -Latex,auto,node distance =2 cm and 2 cm,semithick,
    state/.style ={ellipse, draw, minimum width = 0.7 cm},
    point/.style = {circle, draw, inner sep=0.04cm,fill,node contents={}},
    bidirected/.style={Latex-Latex,dashed},
    el/.style = {inner sep=2pt, align=left, sloped}
}

\usepackage{color}

\usepackage{soul}

\newcommand\Span{\textup{span}}

\begin{document}

\def\spacingset#1{\renewcommand{\baselinestretch}%
{#1}\small\normalsize} \spacingset{1}


\if0\blind
{
  \title{\bf Spectrally Deconfounded Random Forests}
  \author{Markus Ulmer, Cyrill Scheidegger, and Peter Bühlmann\\
    Seminar for Statistics, ETH Zürich, Switzerland}
  \maketitle
} \fi

\if1\blind
{
  \bigskip
  \bigskip
  \bigskip
  \begin{center}
    {\LARGE\bf Spectrally Deconfounded Random Forests}
\end{center}
  \medskip
} \fi

\bigskip
\begin{abstract}
We introduce a modification of Random Forests to estimate functions when unobserved confounding variables are present. The technique is tailored for high-dimensional settings with many observed covariates. We employ spectral deconfounding techniques to minimize a deconfounded version of the least squares objective, resulting in the Spectrally Deconfounded Random Forests (\emph{SDForests}). We demonstrate how the omitted variable bias in estimating a direct effect approaches zero, assuming dense confounding and high-dimensional data. We compare the performance of \emph{SDForests} with that of classical Random Forests in a simulation study and a semi-synthetic setting using single-cell gene expression data. Empirical results suggest that \emph{SDForests} outperform classical Random Forests in estimating the direct regression function, even if the theoretical assumptions are not perfectly met, and that \emph{SDForests} and classical Random Forests have comparable performance in the non-confounded case. We provide an R-Package for \emph{SDForest}, and supplementary materials for this article are available online.
\end{abstract}

\noindent%
\vfill

\newpage

\section{Introduction}
\label{sec:intro}

Random Forests \citep{Breiman2001RandomForests} and their variations, such as Random Survival Forests \citep{Hothorn2005SurvivalEnsembles,Taylor2011RandomForests}, Quantile Regression Forests \citep{Meinshausen2006QuantileForests}, or distributional versions of Random Forests \citep{Hothorn2021PredictiveForests, Cevid2022DistributionalRegression} are successfully applied to a wide range of datasets. In many cases of observational data, however, problems with "omitted variable bias" \citep{Cinelli2020MakingBias,Wilms2021OmittedRelationships} arise. This means that a bias is induced in estimating relationships using standard Random Forest versions when covariates that correlate with other covariates and the response are not observed and included. 

In the setting of causality, this can be viewed as a confounded causal relationship with unobserved confounders \citep{Pearl2009Causality:Inference, Peters2016CausalIntervals}. A popular approach to deal with unobserved confounding is to use instrumental variables (IV) regression techniques \citep{Bowden1990InstrumentalVariables, Angrist1996IdentificationVariables, Stock2003Retrospectives:Regression}. Finding enough strong and valid instrumental variables can be challenging, especially if many covariates with potential effects on the response are observed, since the number of instruments must be as large as the number of effective covariates. 

Another possible way of reducing the hidden confounding bias, which we will adopt here, is to make some kind of "dense confounding effect" assumption, meaning that the non-observed factors or confounders affect most of the covariates. Then, a standard approach is to estimate the hidden confounding using methods from high-dimensional factor analysis \citep{Bai2003InferentialDimensions} and explicitly adjust for them, see for example \cite{
Leek2007CapturingAnalysis, Gagnon-Bartsch2012UsingData, Fan2024AreAdequate} for approaches in this direction. Instead of estimating the latent factors explicitly, one can adjust for them implicitly using spectral transformations \citep{Cevid2020SpectralModels}. Applying such spectral transformations, especially the trim transform, which we introduce later, does not require any tuning and is computationally very fast since it is a simple and explicit function of the singular value decomposition of the design matrix. 

\subsection{Our Contribution}
Our contribution is a Random Forest algorithm that is able to address, at least partially, the problem of hidden confounding. Our proposal combines the great advantages and flexibility of standard Random Forests with spectral deconfounding techniques for addressing bias from unobserved factors or confounding. The latter was originally proposed for linear models by \cite{Cevid2020SpectralModels, Guo2022DoublyConfounding} and was further developed for nonlinear additive models by \cite{Scheidegger2025SpectralModels}. The application of spectral deconfounding techniques for Random Forests is novel.

We develop a new algorithm with the R-package \if0\blind{\emph{SDModels} \citep{Ulmer2025SDModels:Models}} \fi \if1\blind{\emph{blinded}} \fi and show its performance in estimating the direct and unconfounded relationship between the observed covariates and the response. We demonstrate that in the presence of hidden confounding, our method, the Spectrally Deconfounded Random Forest (\emph{SDForest}), outperforms the standard Random Forests in many aspects. If no latent factor or confounding exists, our \emph{SDForests} and standard Random Forests perform similarly, perhaps with a minimal edge in favor of the classical algorithm. Thus, if one is unsure whether hidden confounding is present, there is much to gain but not much to lose.

\subsection{Notation}

We denote the largest, the smallest, and $i$-th (non-zero) singular value of any rectangular matrix $A$ by $\lambda_{\max}(A)$, $\lambda_{\min}(A)$ and $\lambda_i(A)$ respectively. The condition number is defined as $\textup{cond}(A) \coloneqq \frac{\lambda_{\max}(A)}{\lambda_{\min}(A)}$.
Let $\{r_n\}_{n = 1}^\infty$ and $\{k_n\}_{n = 1}^\infty$ be positive constants. We use the notation $k_n \coloneqq \Omega(r_n)$ if $\frac{r_n}{k_n} = \mathcal{O}(1)$, i.e., if $k_n$ has asymptotically at least the same rate as $r_n$ and $k_n \asymp r_n$ if $k_n$ and $r_n$ have asymptotically the same rate. We write $r_n \ll  k_n$ if $\frac{r_n}{k_n} = o(1)$.

\section{Confounding Model}

Throughout this work, we assume the confounding model, written in terms of structural equations
\begin{equation} \label{eq:confounding_model}
\begin{split}
    & X \leftarrow \Gamma^T H + E \\
    & Y \leftarrow f^0(X) + \delta^T H + \nu.
\end{split}
\end{equation}
Here, $X \in \mathbb{R}^{p}$ are the predictors, $Y \in \mathbb{R}$ is the response, and $H \in \mathbb{R}^{q}$ are unobserved hidden confounding factors. We assume that the confounder influences $X$ with a linear effect $\Gamma \in \mathbb R^{q\times p}$ and $Y$ with a linear effect $\delta \in \mathbb R^q$ (see Appendix C for a note on non-linear confounding); without loss of generality, we can assume $\mathrm{Cov}(H) = I$. Furthermore, $\nu$ is a random variable with mean zero and variance $\sigma_{\nu}^2$, and $E$ is a random vector with mean zero and covariance $\Sigma_{E}$, and $E$ and $H$ are uncorrelated. The error term $E$ can be viewed as the unconfounded predictor if $\Gamma$ equals zero. Finally, $f^0 \in \mathcal{F}$, where $\mathcal{F}$ is some class of functions from $\mathbb{R}^p$ to $\mathbb R$. The function $f^0$ encodes the direct causal relationship of interest, describing the causal relation of $X$ on $Y$. The described model is visualized as a graph in Figure \ref{fig:causal_graph}.

\begin{figure}[hbt!]
    \centering
    \resizebox{0.35\textwidth}{!}{%
    \begin{tikzpicture}
    \node[state] (1) {$X$};
    \node[state] (2) [right =of 1] {$Y$};
    \node (3) at (1.4,1.5) [label=above:$H$,point];
    
    \path (1) edge node[above] {$f^0$} (2);
    \path (3) edge node[above left] {$\Gamma$} (1);
    \path (3) edge node[above right] {$\delta$} (2);
    \end{tikzpicture}
    }%
    \caption{Confounding model \eqref{eq:confounding_model}, with hidden confounder $H$ affecting $X$ and $Y$ linearly. The function $f^0(X)$ encodes 
    the direct effect of $X$ on $Y$.}
    \label{fig:causal_graph}
\end{figure}

\section{Generic Methodology}
\label{sec:theory}

We assume that we observe $n$ i.i.d. observations from $X$ and $Y$ generated by model \eqref{eq:confounding_model}. We concatenate them row-wise into the design matrix $\mathbf X\in \mathbb R^{n\times p}$ and the vector of responses $\mathbf Y\in \mathbb R^n$. Classical regression methods ignoring the confounding would estimate $f^0$ by minimizing the least squares objective $\hat f \coloneqq \text{arg\ min}_{f\in \mathcal F}\|\mathbf Y -f(\mathbf X)\|_2^2$, or a regularized version thereof, for a suitable class $\mathcal F$ of functions. This yields an estimate of $\text{arg\ min}_{f\in \mathcal F}\mathbb E[(Y-f(X))^2]=\mathbb E[Y|X] =f^0(X) + \delta^T\mathbb E[H|X]$, which is a biased estimate for $f^0$ in model \eqref{eq:confounding_model}. We apply a spectral transformation to the least squares objective to remove this confounding bias. Let $Q\in \mathbb R^{n\times n}$ be a transformation matrix that depends on the data $\mathbf X$. Examples are the trim-transform and the PCA adjustment \citep{Cevid2020SpectralModels, Guo2022DoublyConfounding, Scheidegger2025SpectralModels}. We use the trim-transform in our empirical results in Section \ref{sec:empirical}, which limits all the singular values of $\mathbf X$ to some constant $\tau$, but the methodology can be applied using other spectral transformations. Let $\mathbf{X} = UDV^T$ be the singular value decomposition of $\mathbf{X}$, where $U \in \mathbb R^{n \times r}$, $D \in \mathbb R^{r \times r}$, $V \in \mathbb R^{p \times r}$, where $r \coloneqq \min(n, p)$ is the rank of $\mathbf{X}$. The trim transform $Q$ is then defined as
\begin{equation} \label{eq:Q}
    Q \coloneqq U \tilde{D} U^T
\end{equation}
\begin{equation*}
    \tilde{D} \coloneqq 
    \begin{bmatrix} 
    \tilde{d_1}/d_1 & 0 & \cdots & 0 \\
    0 & \tilde{d_2}/d_2 & \cdots & 0 \\
    \vdots & \vdots & \ddots & \vdots \\
    0 & 0 & \cdots & \tilde{d_r}/d_r
    \end{bmatrix}
\end{equation*}
\begin{equation*}
    \tilde{d_i} \coloneqq \min(d_i, \tau)
\end{equation*}
with $\tau$ being the median singular value of $\mathbf{X}$, as recommended by \cite{Cevid2020SpectralModels} (see Appendix B for a visualization). Trimming the first few singular values results in the reduction of the loss in the direction of the first few principal components of $\mathbf{X}$ and in the confounding model \eqref{eq:confounding_model}, this is also the direction containing most of the confounding effects. Thus, reducing this part of the loss results in the reduction of the confounding bias. At the same time, it is very unlikely that the true sparse function $f^0(X)$ lies in the direction of the first few principal components unless there is an artificial relation between $f^0(.)$ and the covariance matrix of $X$. We will, therefore, minimize a spectrally transformed version of the mean squared error that we refer to as the spectral objective:

\begin{equation} \label{eq:spectral_objective}
    \min_{f \in \mathcal{F}} \frac{\|Q(\mathbf{Y} - f(\mathbf{X}))\|_2^2}{n}.
\end{equation}

\begin{figure}[hbt!]
  \centering
  \includegraphics[width=1\textwidth]{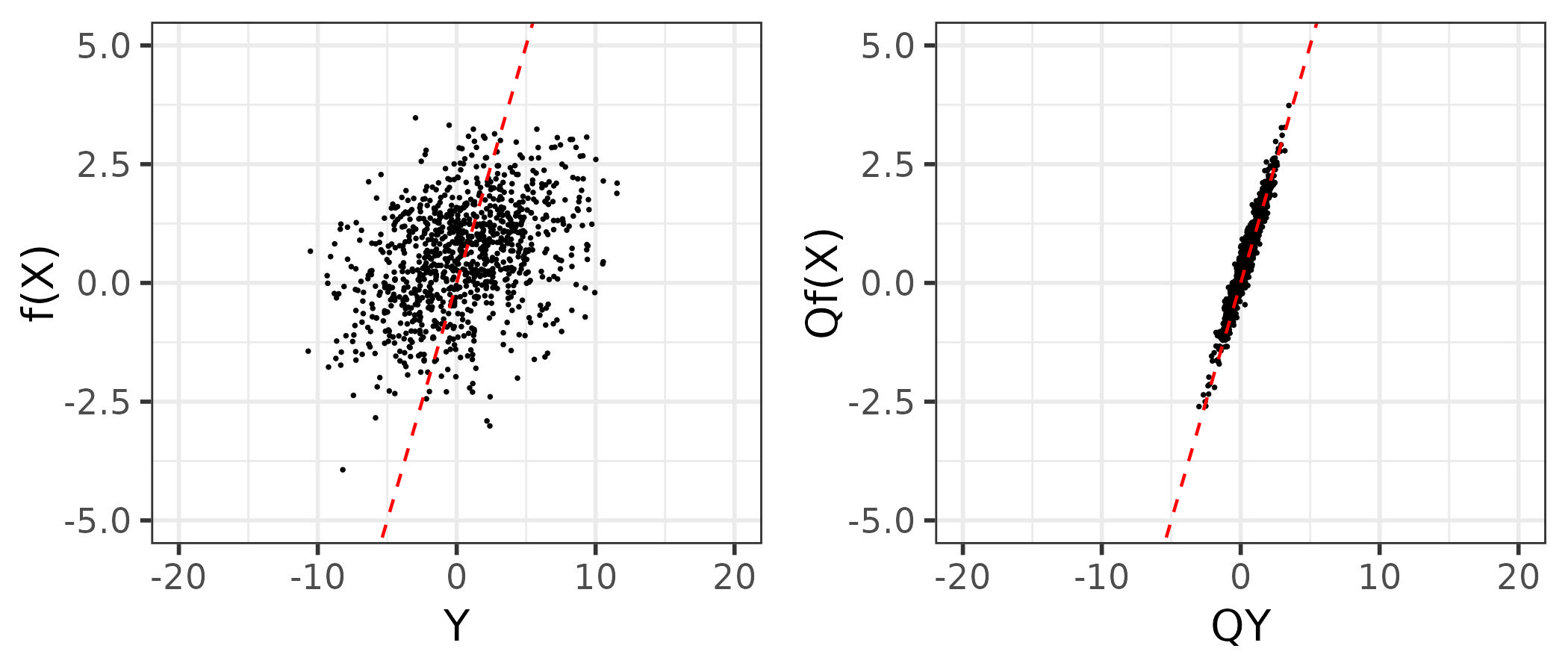}
  \caption{A random realization according to the confounding model \eqref{eq:confounding_model} with non-linear $f^0$ as in \eqref{eq:confounding_model_rf} and with the same parameter as in Section \ref{sec:empirical}. On the left, we show $f^0(\mathbf{X})$ against $\mathbf{Y}$; on the right, the spectrally transformed versions are shown against each other, that is, $Qf^0(\mathbf{X})$ versus $Q\mathbf{Y}$. In both visualizations, the line with a slope of one, corresponding to equality, is shown as a dashed line.}
  \label{fig:pt}
\end{figure}

Theorem \ref{t:objective_prime} below shows that if the confounding follows some assumptions and we have a spectral transformation, we optimize in the limit essentially $\min_{f \in \mathcal{F}} \|Q(f^0(\mathbf{X}) - f(\mathbf{X}) + \nu)\|_2^2 / n$ with $\nu$ as in model \eqref{eq:confounding_model} and being independent of $X$.
This means that we asymptotically minimize a transformed least squares objective without confounding rather than the usual least squares objective with confounding. Figure \ref{fig:pt} shows how the spectral transformation changes the correlation between $\mathbf{Y}$ and $f^0(\mathbf{X})$ and that we can use $Q\mathbf{Y}$ as an approximation for $Qf^0(\mathbf{X})$.

\subsection{Assumptions and Technical Motivation}\label{sec:assumptions}

The spectral deconfounding methodology \citep{Cevid2020SpectralModels} relies on a set of assumptions that are also crucial for our \emph{SDForests}. We review these assumptions in the following.
\begin{description}
    \item[Model:] The data is generated according to the confounding model \eqref{eq:confounding_model} with
    \begin{equation}\label{eq:CondModel}
        \mathbb E[H]= 0\in \mathbb R^q, \; \mathbb E[H H^T]  = I_q, \; \mathbb E[E]= 0\in \mathbb R^p, \; \mathbb E[H E^T] = 0\in \mathbb R^{p\times q},
    \end{equation}
    i.e., $H$ and $E$ are both centered, and they are uncorrelated.
    The assumption that $H$ has unit covariance matrix is without loss of generality: let $\Sigma_H \coloneqq \mathbb E[H H^T].$ We can then consider the confounding model with $\tilde H \coloneqq \Sigma_H^{-1/2} H$, $\tilde \Gamma \coloneqq \Sigma_H^{1/2} \Gamma$ and $\tilde \delta \coloneqq \Sigma_H^{1/2} \delta$ which satisfies $\mathbb E[\tilde H\tilde H^T]= I_q$.
    
    \item[Dimensions:] We will see in Theorem \ref{t:objective_prime} below that the confounding effect goes to zero as $\min (n, p)$ grows. Hence, we need to assume that $p$ increases to infinity with $n$. Moreover, we need the number $q$ of confounders to be low-dimensional, i.e., $q\ll \min(n, p)$.
    
    \item[Covariance of $E$:] It is essential that the covariance $\Sigma_E \coloneqq \mathbb E[E E^T]\in \mathbb R^{p\times p}$ of the unconfounded part $E$ of $X$ is sufficiently well-behaved. If, for example, $E$ itself had a factor structure, it would be difficult to separate the confounding $\Gamma^T H$ from the factor structure in $E$. This well-behavedness assumption is formalized by
    \begin{equation}\label{eq:CondSigmaE}
        \textup{cond}(\Sigma_E) = \mathcal O(1)\text{ and } \lambda_{\max}(\Sigma_E) = \mathcal O(1).
    \end{equation}
    A simple example where \eqref{eq:CondSigmaE} is satisfied is when the components of $E$ are uncorrelated and of the same order, i.e., $\Sigma_E = \textup{diag}(\sigma_1^2, \ldots, \sigma_p^2)$ with $\max_{i,j = 1,\ldots, p}\sigma_i^2/\sigma_j^2\leq C_1<\infty$ and $\max_{i= 1,\ldots, p}\sigma_i^2 \leq C_2 <\infty$ for some constants $C_1, C_2 >0$ independent of $p$. However, more general covariance structures are possible.
    
    \item[Dense confounding:] The dense confounding assumption intuitively means that each component of $H$ affects many components of $X$ and hence is a property of the matrix $\Gamma$. More formally, it is a statement on how large the minimal singular value of $\Gamma$ should be. For Theorem \ref{t:objective_prime} below, we will work under the assumption that 
    \begin{equation}\label{eq:DenseConfounding}
        \lambda_{\min}(\Gamma) = \Omega(\sqrt p),
    \end{equation}
    although weaker assumptions are possible \citep{Guo2022DoublyConfounding, Scheidegger2025SpectralModels}. Equation \eqref{eq:DenseConfounding} is, for example, satisfied with high probability if $q/p\to 0$ and either the rows or columns of $\Gamma$ are sampled as i.i.d. sub-Gaussian random vectors with mean zero and covariance $\Sigma_\Gamma$ with $\lambda_{\min}(\Sigma_\Gamma)$ bounded away from zero (see Lemma 6 in \cite{Cevid2020SpectralModels}).

    \item[Spectral transformation: ] The spectral transformation $Q$ defined in \eqref{eq:Q} (trim transform) satisfies
    \begin{equation}\label{eq:QProperty}
        \lambda_{\max}(Q\mathbf X) = \mathcal  O_P\left(\sqrt{\max(n,p)}\right).
    \end{equation}
    In \cite{Guo2022DoublyConfounding}, \eqref{eq:QProperty} is verified for the case where $E$ is a sub-Gaussian random vector and $\lambda_{\max}(\Sigma_E) = \mathcal O(1).$
\end{description}

The following theorem, which is essentially a compilation of results from \cite{Cevid2020SpectralModels} and \cite{Guo2022DoublyConfounding}, serves as a motivation to construct \emph{SDForests} based on the spectral objective \eqref{eq:spectral_objective}.

\begin{theorem} \label{t:objective_prime}
    Assume the confounding model (\ref{eq:confounding_model}) and assume that the conditions \eqref{eq:CondModel}, \eqref{eq:CondSigmaE}, \eqref{eq:DenseConfounding}, and \eqref{eq:QProperty} hold. Then, it holds that
    
    \begin{equation*}
        \frac{\|Q(\mathbf{Y} - f(\mathbf{X}))\|_2}{\sqrt n} = \frac{\|Q(f^0(\mathbf{X}) - f(\mathbf{X}) + \nu)\|_2}{\sqrt{n}} + R_n
    \end{equation*}

    where $R_n = \mathcal{O}_{\mathbb P}\left( \frac{\|\delta\|_2}{\min(\sqrt{n}, \sqrt{p})}\right)$. 
\end{theorem}
In particular, if $\|\delta\|_2^2\ll \min(n, p)$, we have that $R_n = o_P(1)$. The condition $\|\delta\|_2^2\ll \min(n, p)$ holds for example if $q\ll \min(n, p)$ and all the entries of $\delta\in \mathbb R^q$ are bounded.

Theorem \ref{t:objective_prime} justifies the minimization of the spectral objective \eqref{eq:spectral_objective} to remove the confounding bias. Additionally, to effectively estimate $f^0(\cdot)$ in the high-dimensional setting, it seems crucial that $f^0(\cdot)$ is sparse. In fact, for the theory in the linear and the additive case \citep{Cevid2020SpectralModels, Guo2022DoublyConfounding, Scheidegger2025SpectralModels}, sparsity is a central requirement for consistency of spectral deconfounding, so we expect sparsity to also play an important role here.

\section{\emph{SDForest} Algorithm}
\label{sec:algorithm}

In principle, our algorithm works similarly to the original \emph{CART} algorithm \citep{Breiman2017ClassificationTrees} and Random Forests \citep{Breiman2001RandomForests}. \emph{CART} minimizes the mean squared error by greedily dividing the space $\mathbb{R}^p$ into rectangular parts. Each region then has a response level resulting in a function as in Equation \eqref{eq:confounding_model_tree} below. \emph{CART} starts with a single region and then searches in all the variables and along their support for the split that minimizes the mean squared error. This process is repeated for both subsequent regions and continues in the same manner. The mean squared error has the nice property that the loss decomposes, meaning that for the next optimal split in a region, \emph{CART} only needs to look at the samples belonging to this region. The main difference between our algorithm and \emph{CART} is that we minimize the spectral objective \eqref{eq:spectral_objective} instead of the classical mean squared error. This results in additional challenges as the spectral objective is no longer decomposable.

\subsection{Spectrally Deconfounded Tree}
Assume the confounding model (\ref{eq:confounding_model}) with $\mathcal{F}$ being the function class of step functions, e.g., 

\begin{equation}
    f^0(X) \coloneqq \sum_{m = 1}^M \mathds{1}_{\{X \in R_m\}} c_m,
    \label{eq:confounding_model_tree}
\end{equation}
where $(R_m)_{m=1}^M$ are regions dividing the space of $\mathbb R^p$ into $M$ disjoint rectangular parts. Each region has a response level $c_m \in \mathbb R$. We can write the sample version as $f^0(\mathbf{X}) = \mathcal{P} c$ where $\mathcal{P} \in \{0, 1\}^{n \times M}$ is an indicator matrix encoding to which region an observation belongs, i.e. $\mathcal P_{i,m}=1$ if the $i$th row of $\mathbf X$ belongs to $R_m$ and $\mathcal P_{i,m}=0$ otherwise. We refer to $\mathcal{P}$ also as a partition, slightly abusing terminology. The vector $c=(c_1, \ldots, c_M)\in \mathbb R^M$ contains the levels corresponding to the different regions. We can estimate $\hat{f}$ in the spirit of \eqref{eq:spectral_objective} with

\begin{equation} \label{eq:tree}
    (\hat{\mathcal{P}}, \hat{c}) \coloneqq \text{arg\ min}_{\mathcal{P}' \in \mathfrak{P}, c' \in \mathbb R^ {M}} \frac{\|Q(\mathbf{Y} - \mathcal{P'} c')\|_2^2}{n},
\end{equation}
where $\mathfrak{P} \in \{0, 1\}^{n \times M}$ and $\mathfrak{P}$  has to result from a repeated splitting of the space of $\mathbb{R}^p$ into rectangular regions. Repeated splitting means that $\mathfrak{P}$ can be represented by a tree structure. Each branching in the tree corresponds to the splitting of $\mathbb {R}^p$ at a variable $j$ and a split point $s$. 

$M$ is fixed here, but has to be estimated in practice. The estimator $\hat{M}$ for $M$ must be regularized to prevent overfitting. If the partition $\mathcal{P}$ is known, the spectral objective \eqref{eq:spectral_objective} becomes

\begin{equation} \label{eq:c_hat}
    \begin{split}
        \hat{c} & = \text{arg\ min}_{c' \in \mathbb R^M} \frac{\|Q(\mathbf{Y} - \mathcal{P} c')\|_2^2}{n} \\
        & = \text{arg\ min}_{c' \in \mathbb R^M} \frac{\|\mathbf{\tilde{Y}} - \mathcal{\tilde{P}} c'\|_2^2}{n},
    \end{split}
\end{equation}
where $\mathbf{\tilde{Y}} \coloneqq Q\mathbf{Y}$ and $\mathcal{\tilde{P}} \coloneqq Q\mathcal{P}$.
This is a linear regression problem, and we compute $\hat{c}$ with
\begin{equation}
    \hat{c} = (\mathcal{\tilde{P}}^T \mathcal{\tilde{P}})^{-1} \mathcal{\tilde{P}}^T \mathbf{\tilde{Y}}.
\end{equation}

In the spirit of \emph{CART}, we propose Algorithm \ref{alg:SDTree} to find a tree representing $\hat{\mathcal{P}}$ and $ \hat{c}$. While growing the tree, we try to find the next best split. The next best split needs to be chosen among all the current leaves, all variables, and somewhere in $\mathbb{R}$. To calculate how much each split reduces the spectral loss \eqref{eq:spectral_objective}, we employ a subroutine described in the next section. Using these loss decreases, we select the split that yields the maximum reduction in training loss. 

In a subsequent step, the splits in the unused leaves resulting in the maximal loss decrease may no longer be the same. This is due to the induced dependency by the spectral transformation. However, to save computation time, we do not recalculate the optimal split and loss decrease for all leaves, but instead calculate optimal splits and loss decreases only for newly created leaves. For the old leaves, we reuse the previously calculated ones as approximations. We argue that the change is minor and still yields reasonable splits. (In the software, both options are available.)  Appendix A shows a comparison of the performance of this approximation. Note that we still 
re-estimate $\hat{c}$ after each iteration.

To avoid overfitting, a split is only carried out if it reduces the spectral loss \eqref{eq:spectral_objective} enough. The strength of this regularization is controlled by the cost-complexity parameter $\mathsf{cp}$. Only splits that reduce the loss at least as much as $\mathsf{cp}$ times the initial loss, i.e., $\mathsf{cp} \times \frac{\|Q(\mathbf{Y} - \bar{\mathbf{Y}})\|_2^2}{n}$.

\subsection{Subroutine to Evaluate a Split} 
\label{sec:subroutine}
This section explains how we can search for and evaluate potential splits more efficiently.
In the $M$th iteration of Algorithm \ref{alg:SDTree}, we have the indicator matrix $\hat {\mathcal P} = \hat {\mathcal P}^M\in \{0,1\}^{n\times M}$ with entries $\hat{\mathcal P}^M_{i, m} = 1$ if and only if the $i$th observation lies in region $m$. 
We encode a candidate split in the region $b$ of the partition using covariate $j$ at the splitting point $s$ by $e\in \{0,1\}^n$ where $e$ must be in the support of the $b$th column of $\hat {\mathcal P}^M$ (i.e. the indices of $1$s in $e$ are a subset of the indices of $1$s of the $b$th column of $\hat {\mathcal P}^M$)) and the entries depend on $j$ and $s$. A candidate split results in a candidate partition $\mathcal P^{M+1}(e)$ with $M-1$ columns equal to columns $\hat {\mathcal P}^M$, the $b$th column equal to the $b$th column of $\hat P^M$ minus $e$, and an additional column equal to $e$.

In lines \ref{algl:B} to \ref{algl:a} of Algorithm \ref{alg:SDTree}, we seek to find the optimal $e$ among a large number of candidate splits such that the spectral objective $\|QY-Q\mathcal P^{M+1}(e)\hat c\|_2^2$ is minimal, where $\hat c$ is the least squares estimator $\hat c \coloneqq \arg\min_{c\in \mathbb R^{M + 1}}\|QY-Q \mathcal P^{M+1}(e)c\|_2^2 $.
Naively, for every candidate $e$, one would update the indicator matrix $\hat{\mathcal P}$, calculate the corresponding least squares estimator $\hat c$ and plug it in to obtain the loss decrease. Using the following procedure, the decrease in loss can be calculated more efficiently.

Note that $\Span(Q\mathcal P^{M+1}(e)) = \Span(Q\hat{\mathcal P}^M, Q e)$. We set $u_1' \coloneqq Q\cdot(1,\ldots, 1)^T$ and $u_1 \coloneqq u_1'/\|u_1'\|_2$. We proceed by induction and assume that we already have $u_2,\ldots, u_M$ such that  $(u_1,\ldots, u_M)$ form an orthonormal basis for $\Span(Q\hat {\mathcal P}^M)$. Consequently, $\Span(Q\mathcal P^{M+1}(e)) = \Span(Q\hat {\mathcal P}^M, Qe) = \Span(u_1, \ldots, u_M, Qe) = \Span(u_1, \ldots, u_M, u_{M+1}(e))$, where we use orthogonalization to obtain $u_{M+1}(e)$, i.e. $u_{M+1}(e) =u_{M+1}(e)'/\|u_{M+1}(e)'\|_2$ with $u_{M+1}(e)' = Qe - \sum_{l=1}^M (u_l^TQe)u_l = (Q-\sum_{l=1}^M u_l u_l^T Q)e$.

Let $\Pi_{M+1}(e)\in \mathbb R^{n\times n}$ be the orthogonal projection on $\Span(Q\mathcal P^{M+1}(e))=\Span(Q\hat {\mathcal P}^M, Qe) = \Span(u_1,\ldots, u_M, u_{M+1}(e))$. We seek for $e$ that minimizes $\|Q\mathbf Y-Q\mathcal P^{M+1}(e)\hat c\|_2^2=\|(I_n - \Pi_{M+1}(e)) Q\mathbf Y\|_2^2$. Because $(u_1,\ldots, u_M, u_{M+1}(e))$ is an orthonormal set, it follows that 
$$\|(I_n - \Pi_{M+1}(e)) Q\mathbf Y\|_2^2=\|Q\mathbf Y\|_2^2- \|\Pi_{M+1}(e)) Q\mathbf Y\|_2^2=\|Q\mathbf Y\|_2^2- \sum_{l=1}^M (u_l^TQ\mathbf Y)^2- (u_{M+1}(e)^T Q\mathbf Y)^2.$$
To find the optimal split $e$, it suffices to maximize $\alpha(e) \coloneqq (u_{M+1}(e)^T Q\mathbf Y)^2$ among the candidate splits. Once the optimal split $e^\ast$ is found, one can define $u_{M+1} \coloneqq u_{M+1}(e^\ast)$ and $\hat{\mathcal P}^{M+1} \coloneqq \mathcal P^{M+1}(e^\ast)$ and proceed with step $M+2$.

\begin{algorithm}
\caption{Spectrally Deconfounded Regression Tree}\label{alg:SDTree}
\begin{algorithmic}[1]
\Inputs{$\mathbf{X} \in \mathbb R^{n \times p}$, $\mathbf{Y} \in \mathbb R^n$, $Q \in \mathbb R^{n \times n}$, $\mathsf{cp} \in \mathbb R$, $M_{max} \in \mathbb{N}$}
\Initialize{$M \gets 1$ \\ 
$\hat{\mathcal{P}} \gets (1,\ldots, 1)^T \in \mathbb R^{n\times 1}$ \\ 
$\tilde{\hat{\mathcal{P}}} \gets Q\hat{\mathcal{P}}$ \\
$\mathbf{\tilde{Y}} \gets Q\mathbf{Y}$\\
$u \gets \hat{\mathcal{P}} / \|\hat{\mathcal{P}}\|_2$\\
$Q^d \gets Q - u u^T Q$ \\
$\hat{c} \in \mathbb R^M \gets \text{arg\ min}_{c' \in \mathbb R^m} \|\mathbf{\tilde{Y}} - \tilde{\hat{\mathcal{P}}} c'\|_2^2 / n$ \\ 
$l^{init} \gets l^{temp} \gets \|\mathbf{\tilde{Y}} - \tilde{\hat{\mathcal{P}}} \hat{c}\|_2^2 / n$ \\
$l^{dec} \in \mathbb R^M \gets 0$ \\
$\mathcal{B} \gets 1$}

\For{$M = 1$ to $M_{max}$} \label{algl:M}
    \For{$b$ in $\mathcal{B}$}   \label{algl:B} \Comment{subroutine}
        \For{$(j, s)$ in potential splits in region $b$}
            \State $e_{b, j, s} \in \{0,1\}^n \gets$ indices of samples belonging to the new partition
            \State $u \gets Q^d e_{b, j, s} / \|Q^d e_{b, j, s}\|_2$
            \State $\alpha_{b, j, s} \gets (u^T \mathbf{\tilde{Y}})^2$
        \EndFor
    \EndFor \label{algl:a}
    \State $(b^*, j^*, s^*) \gets \text{arg\ max }{\alpha_{b, j, s}}$ \Comment{optimal split over $b \in \{1, \ldots, M\}$} \label{algl:besta}
    \State $u \gets Q^d e_{b^*, j^*, s^*} / \|Q^d e_{b^*, j^*, s^*}\|_2$
    \State $Q^d \gets Q^d - uu^TQ$
    \State $\hat{\mathcal{P}}^* \gets$ splitting $\hat{\mathcal{P}}$ at $(b^*, j^*, s^*)$ \Comment{resulting in $\hat{\mathcal{P}}^* \in \mathbb{R}^{n \times (M+1)}$}
    \State $\hat{c}^* \gets \text{arg\ min}_{c' \in \mathbb R^M} \|\mathbf{\tilde{Y}} - \tilde{\hat{\mathcal{P}}}^* c'\|_2^2 / n$ \label{algl:c}
    \State $l^* \gets \|\mathbf{\tilde{Y}} - \tilde{\hat{\mathcal{P}}}^*\hat{c}^*\|_2^2 / n$
    \State $d \gets l^{temp} - l^*$ \label{algl:bestd}
    \If{$d > \mathsf{cp} * l^{init}$}
        \State $\hat{\mathcal{P}} \gets \hat{\mathcal{P}}^*$
        \State $\hat{c} \gets \hat{c}$
        \State $l^{temp} \gets l^*$ 
        \State $\mathcal{B} \gets (b^*, M + 1)$ \label{algl:Bnew} \Comment{the new partitions}
        \Else
        \State break
    \EndIf
\EndFor
\end{algorithmic}
\end{algorithm}

\subsection{Spectrally Deconfounded Random Forests}

The next natural step is to utilize spectrally deconfounded regression trees (SDTree) to construct spectrally deconfounded Random Forests (\emph{SDForests}) for estimating arbitrary functions. Random forests have been introduced by \cite{Breiman2001RandomForests} and have been successfully employed in numerous applications. The idea is to combine multiple regression trees into an ensemble to decrease variance and obtain a smoother function. 

Ensembles work best if the different models are independent of each other. To decorrelate the regression trees as much as possible, we employ two mechanisms. The first one is bagging \citep{Breiman1996BaggingPredictors}, where we train each regression tree on an independent bootstrap sample of the observations, i.e., we draw a random sample of size $n$ with replacement from the observations. The second mechanism to decrease the correlation is that only a random subset of the covariates is available for each split. Before each split, we sample $\mathsf{mtry} \leq p$ from all the covariates and choose the one that reduces the loss the most from those. 

It only takes minor changes to Algorithm \ref{alg:SDTree} to build an \emph{SDForest}. In Algorithm \ref{alg:SDTree} before line \ref{algl:besta}, we sample a set $p_\mathsf{mtry}$ of size $\mathsf{mtry}$ from the covariates, and in line \ref{algl:besta}, we only check in this set of randomly sampled covariates for the best split, e.g., $j \in p_\mathsf{mtry}$. This procedure yields the spectrally deconfounded Random Forest, as outlined in Algorithm \ref{alg:SDForest}. We predict with all the trees separately and use the mean over all trees as the Random Forest prediction, resulting in 

\begin{equation}
    \hat{f}(X) \coloneqq \frac{1}{N_{tree}} \sum_{t = 1}^{N_{tree}} SDTree_t(X).
\end{equation}

\begin{algorithm}
\caption{Spectrally Deconfounded Random Forest}\label{alg:SDForest}
\begin{algorithmic}
\Inputs{$\mathbf{X} \in \mathbb R^{n \times p}$, $\mathbf{Y} \in \mathbb R^n$, $Q \in \mathbb R^{n \times n}$, $N_{tree} \in \mathbb{N}$, $\mathsf{mtry} \in [1, p]$}

\For{$t = 1$ to $N_{tree}$}
    \State $X^t \gets$ bootstrap sample of $X$
    \State Find $Q^t$ using $X^t$
    \State $SDtree_t \gets$ SDTree from Algorithm \ref{alg:SDTree} with random set of covariates of size $\mathsf{mtry}$ at each split using $X^t$ and $Q^t$.
\EndFor
\end{algorithmic}
\end{algorithm}

\section{Empirical Results}
\label{sec:empirical}

In this section, we analyze with a simulation study how well \emph{SDForests} estimate a true causal function in comparison to classical Random Forests. Subsection \ref{sec:qual} shows qualitatively how \emph{SDForest} can screen for and estimate a sparse causal effect. The ability to estimate the causal function with respect to different dimensions and on the dense confounding assumption is examined quantitatively in subsection \ref{sec:dim}.

For the simulation study, we simulate data according to the confounding model (\ref{eq:confounding_model}) with a random $f^0$ using the Fourier basis
\begin{equation} \label{eq:confounding_model_rf}
    f^0(X) \coloneqq \sum_{j = 1}^p \mathds{1}_{\{j \in \mathcal J_s\}} \sum_{k = 1}^K (a_{j, k} \cos(0.2 k \cdot x_j) + b_{j, k} \sin(0.2 k \cdot x_j))
\end{equation}
where $\mathcal J_s$ is a random subset of ${1, \ldots, p}$ of size four being the parents of $Y$ (among the $X$ variables). We simulate with $n = 1000$, $p = 500$, and $q = 20$. The entries of $\mathbf{E} \in \mathbb R^{n \times p}$, $\mathbf{H} \in \mathbb R^{n \times q}$, $\delta \in \mathbb R^q$, and $\Gamma \in \mathbb R^{q \times p}$, in the confounding model \eqref{eq:confounding_model}, are sampled i.i.d. from a Gaussian with expectation zero and $\sigma = 1$. The additional noise $\mathbf{\nu} \in \mathbb R^n$ is sampled i.i.d. from a Gaussian with mean zero and $\sigma_{\nu} = 0.1$. For the four causal parents, we sample the coefficients $a_{j, k}$ and $b_{j, k}$ uniformly on $[-1, 1]$ with the number of basis functions fixed at $K = 2$ for the additive function.

We use the \emph{SDForest} with a hundred trees, $\mathsf{mtry} = \lfloor0.5p\rfloor$ and $Q$ as the trim-transform \eqref{eq:Q} \citep{Cevid2020SpectralModels} to estimate the causal function. We compare the results of the \emph{SDForest} to the estimated function by the classical Random Forest using the r package \emph{ranger} \citep{Wright2017Ranger:R}.

\subsection{Qualitative Results}\label{sec:qual}

\begin{figure}[hbt!]
  \centering
  \includegraphics[width=0.71\textwidth]{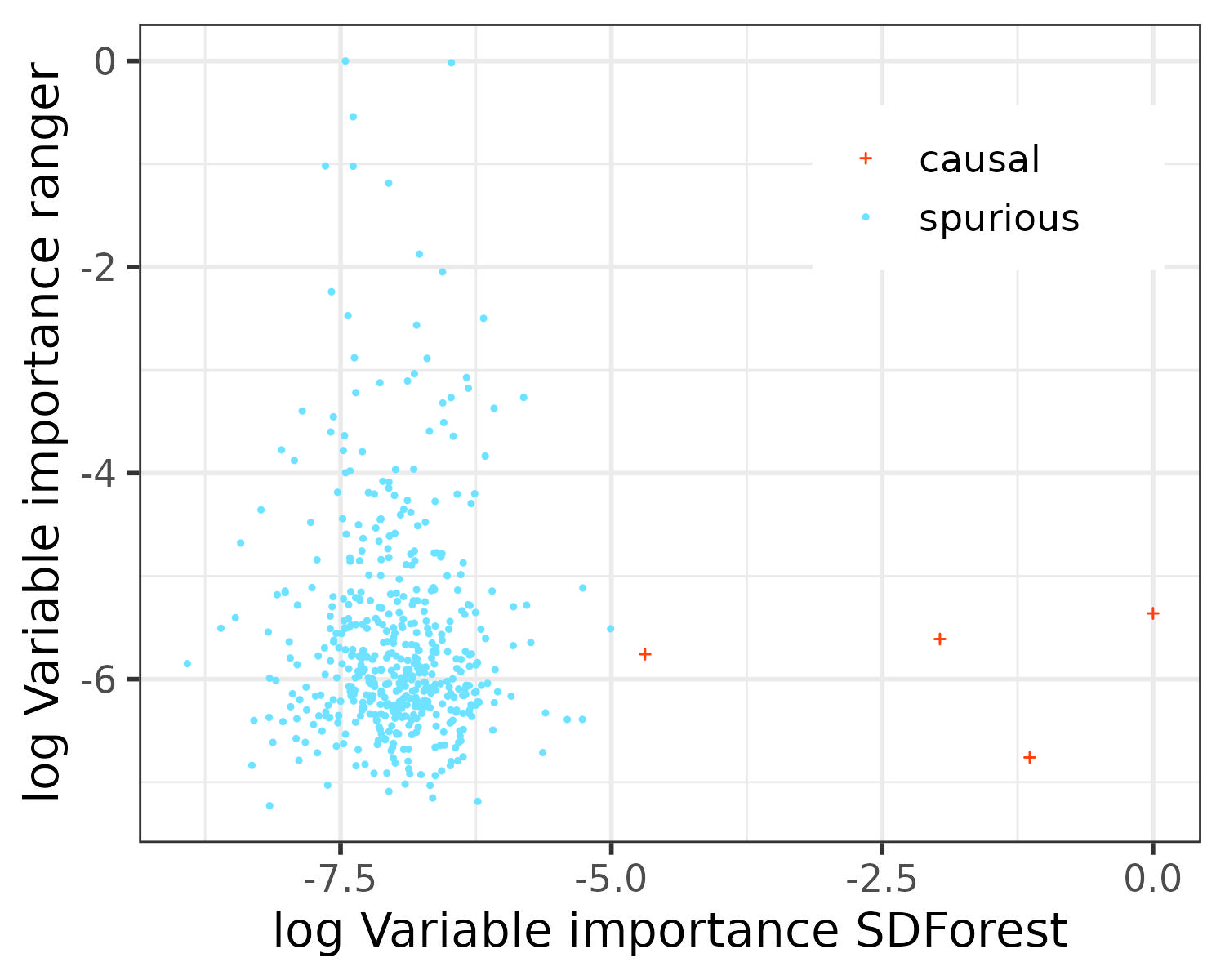}
  \caption{Comparison of variable importance for a realization of model \eqref{eq:confounding_model_rf} between the classical Random Forest estimated by \emph{ranger} and the \emph{SDForest}. The variable importance for both methods is scaled to the interval $[0, 1]$ and log-transformed. The true causal parents of the response $Y$ are marked as crosses.}
  \label{fig:imp}
\end{figure}

In Figure \ref{fig:imp}, we compare the variable importance of the \emph{SDForest} and the classical Random Forest for one simulation run. The variable importance of a covariate is calculated as the sum of the loss decrease resulting from all splits that divide a region using that covariate. The mean of the variable importance across all trees yields the variable importance for the forest. For the \emph{SDForest}, we report the reduction of the spectral loss \eqref{eq:spectral_objective} instead of the MSE. For the \emph{SDForest}, the four most important covariates are also the four true causal parents of $Y$. Three of those also have a clear separation from the remaining 497 covariates. For the classical Random Forest, the variable importance for the true causal parents lies within the bulk of spurious covariates. The most important covariate among the true causal parents is only on rank 102, and we would have to use almost all the covariates to include all the causal parents in the selected set.

\begin{figure}[hbt!]
  \centering
      \includegraphics[width=1\textwidth]{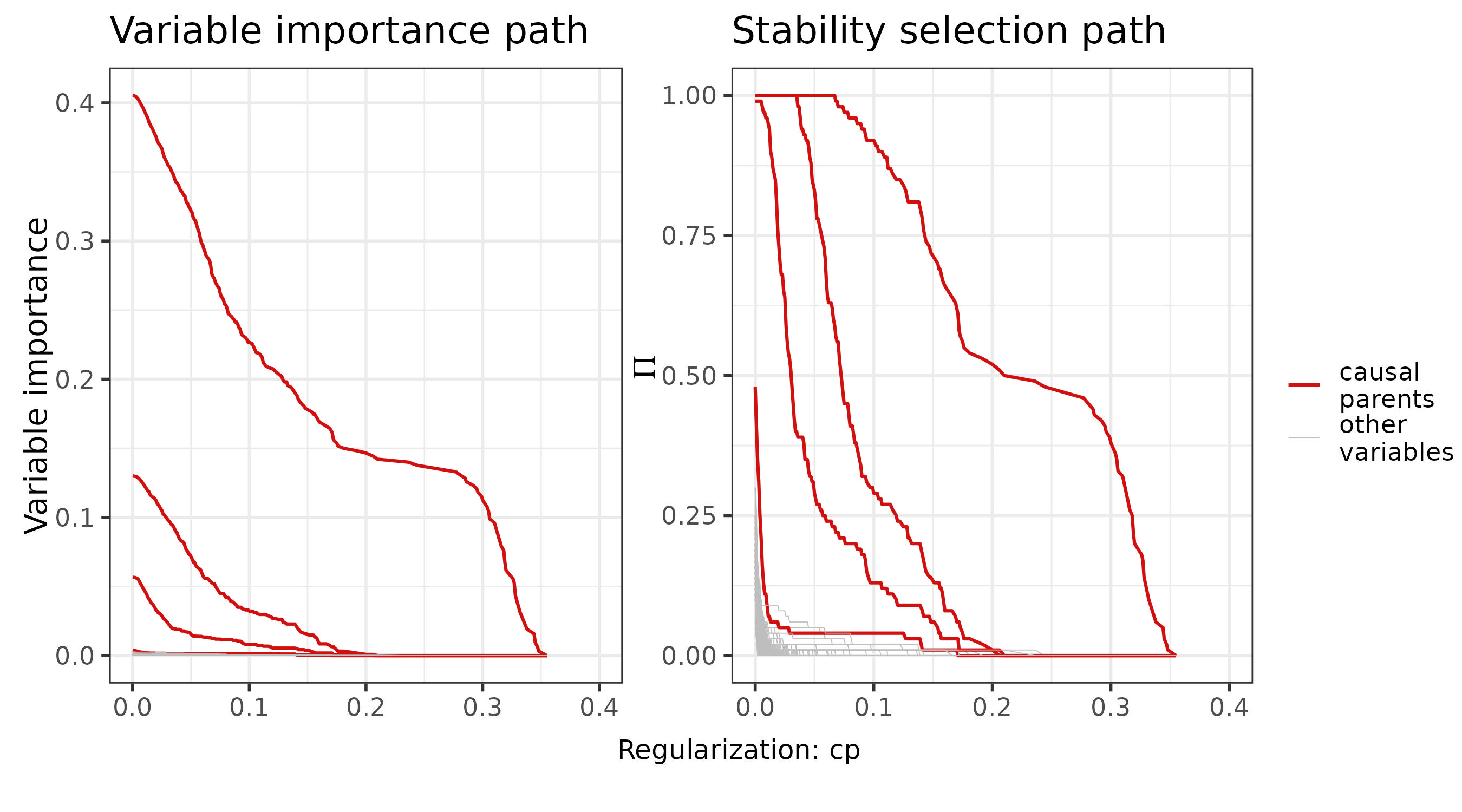}
  \caption{Regularization paths of the \emph{SDForest} estimated on a realization of model \eqref{eq:confounding_model_rf} when varying the cost-complexity parameter $\mathsf{cp}$ resulting in more or less pruned trees. Each curve corresponds to a single covariate. On the left side are the variable importance paths for different strengths of regularization shown. On the right side are the stability selection paths against the strength of regularization shown. $\Pi$ corresponds to the ratio of trees in the forest that use a covariate. The truly causal parents of the response $Y$ correspond to the darker, thicker lines.}
  \label{fig:paths}
\end{figure}

Instead of examining the variable importance of the fully grown trees in the \emph{SDForest}, we can also examine the regularization paths of the covariates. Figure \ref{fig:paths} shows on the left side the variable importance for the \emph{SDForest} against increasing regularization, where one increases $\mathsf{cp}$ subsequently pruning the trees. Here again, three of the causal parents appear. On the right side in Figure \ref{fig:paths}, we show stability selection \citep{Meinshausen2010StabilitySelection}, where $\Pi$ is the ratio of trees that use a particular covariate in the forest given increasing regularization. In the stability selection paths, we also see the fourth causal parent.

In addition to screening for the sparse causal set among a large number of covariates, we can also look at the functional dependence of the response $Y$ on the causal parents. We use partial dependence plots \citep{Friedman2001GreedyMachine} to visualize the partial dependence. The idea is to predict $\hat{f}(\mathbf{X})$ for each observation and vary $\mathbf{X}_j$ over an interval, while keeping the other covariates at their observed values. This yields a different function of $X_j$ for every observation. The mean over all the observations can then be shown as a representative marginal effect of $X_j$ on the response $Y$. 

\begin{figure}[hbt!]
  \centering
  \includegraphics[width=1\textwidth]{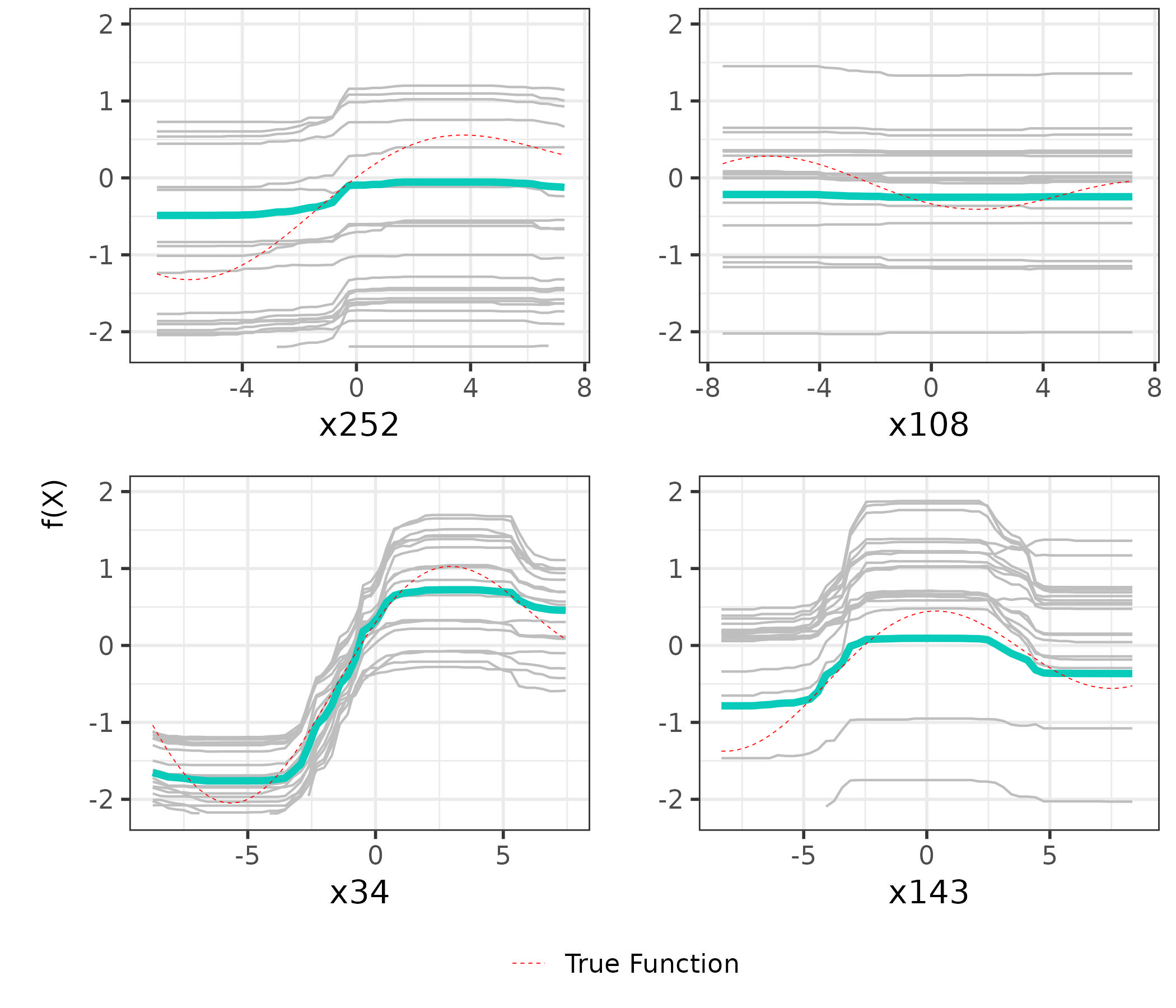}
  \caption{Partial dependence plots of the estimated \emph{SDForest} for the four true causal parents of the response $Y$. The dashed line is the corresponding true partial causal function. The light lines show the observed empirical partial functions for 19 randomly selected observations, and the thick line is the average of all observed partial functions.}
  \label{fig:cond_rf}
\end{figure}

\begin{figure}[hbt!]
  \centering
  \includegraphics[width=0.5\textwidth]{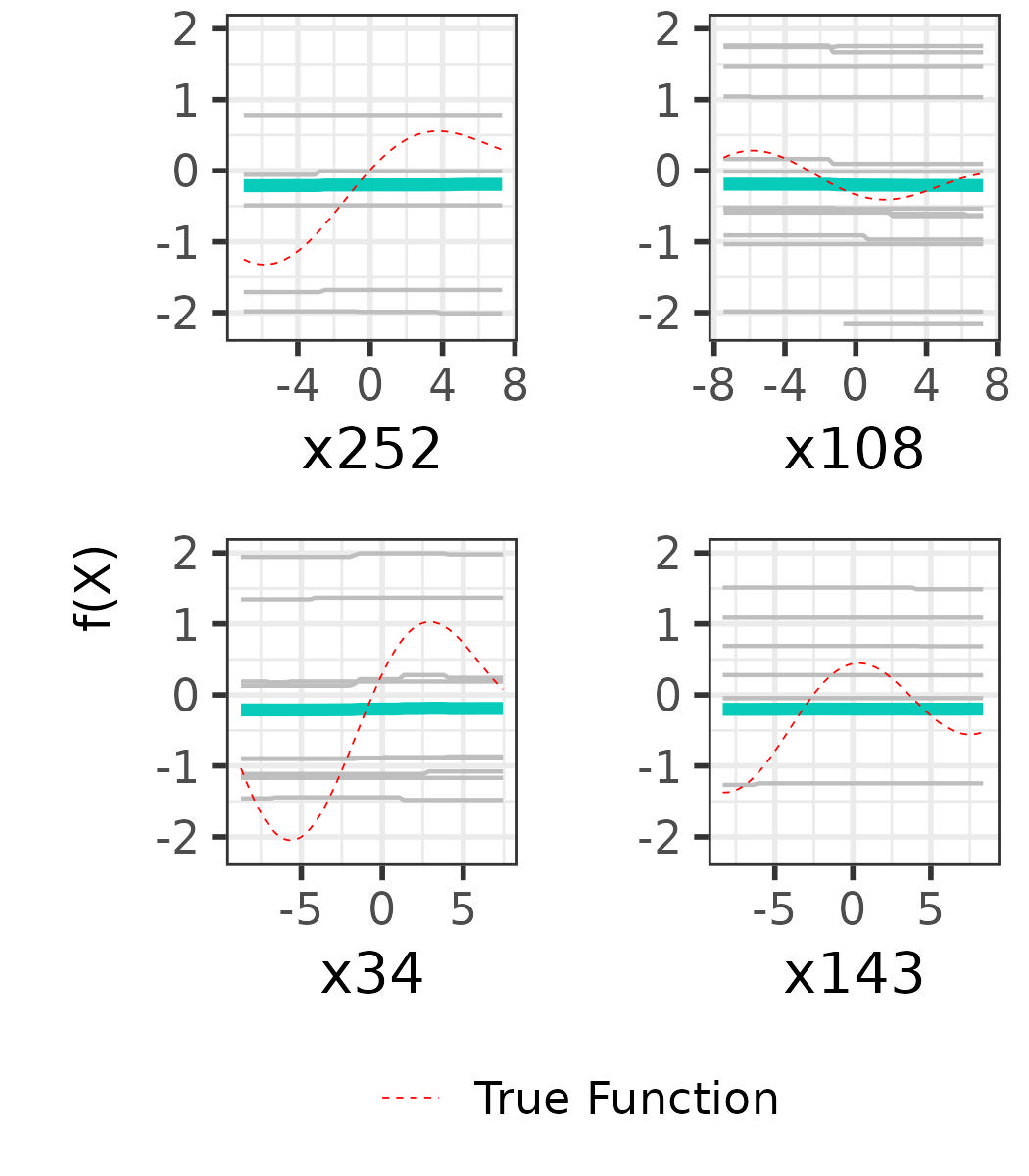}
  \caption{Partial dependence plots as in Figure \ref{fig:cond_rf} but now with the estimated classical Random Forest by \emph{ranger} for the four true causal parents of the response $Y$. The classical Random Forest essentially leads to constant functions, whereas the true dashed lines vary.} 
  \label{fig:cond_r}
\end{figure}

Figure \ref{fig:cond_rf} shows how the \emph{SDForest} estimates the true causal function. Especially for covariate 34, the estimated function closely approximates the true function. Covariate 108, the covariate with only slightly higher importance than the bulk of spurious covariates, has almost a constant influence on $Y$. This shows some limitations of estimating a sparse causal relationship in the presence of hidden confounding and high dimensionality. Estimating a nuanced true function, such as for variable 108, might be difficult due to the additional disturbances of confounding and noise. The classical Random Forest distributes the estimated function among all the spurious covariates and estimates almost a constant function for all four causal parents. The partial dependence plots for the estimated classical Random Forest are shown in Figure \ref{fig:cond_r}.

\subsection{Quantitative Results}\label{sec:dim}

To quantify the performance of \emph{SDForests} in estimating the causal function, we conduct simulation studies that vary different dimensions in the simulation. The default dimensions are $n = 500$, $p = 500$, and $q = 20$, and we randomly draw data according to model \eqref{eq:confounding_model_rf}. Each of those dimensions is varied separately to estimate the dependence of the performance on these different factors. Every experiment is repeated 200 times, where the entire data-generating process is redrawn at random each time. The performance is measured by the mean distance of the true causal function and the estimated function evaluated at 500 test observations $f_{mse} \coloneqq \frac{1}{n_{test}}\sum_{i=1}^{n_{test}}(f^0(x_{test, i})- \hat f(x_{test, i}))^2$.

\begin{figure}[hbt!]
  \centering
  \includegraphics[width=1\textwidth]{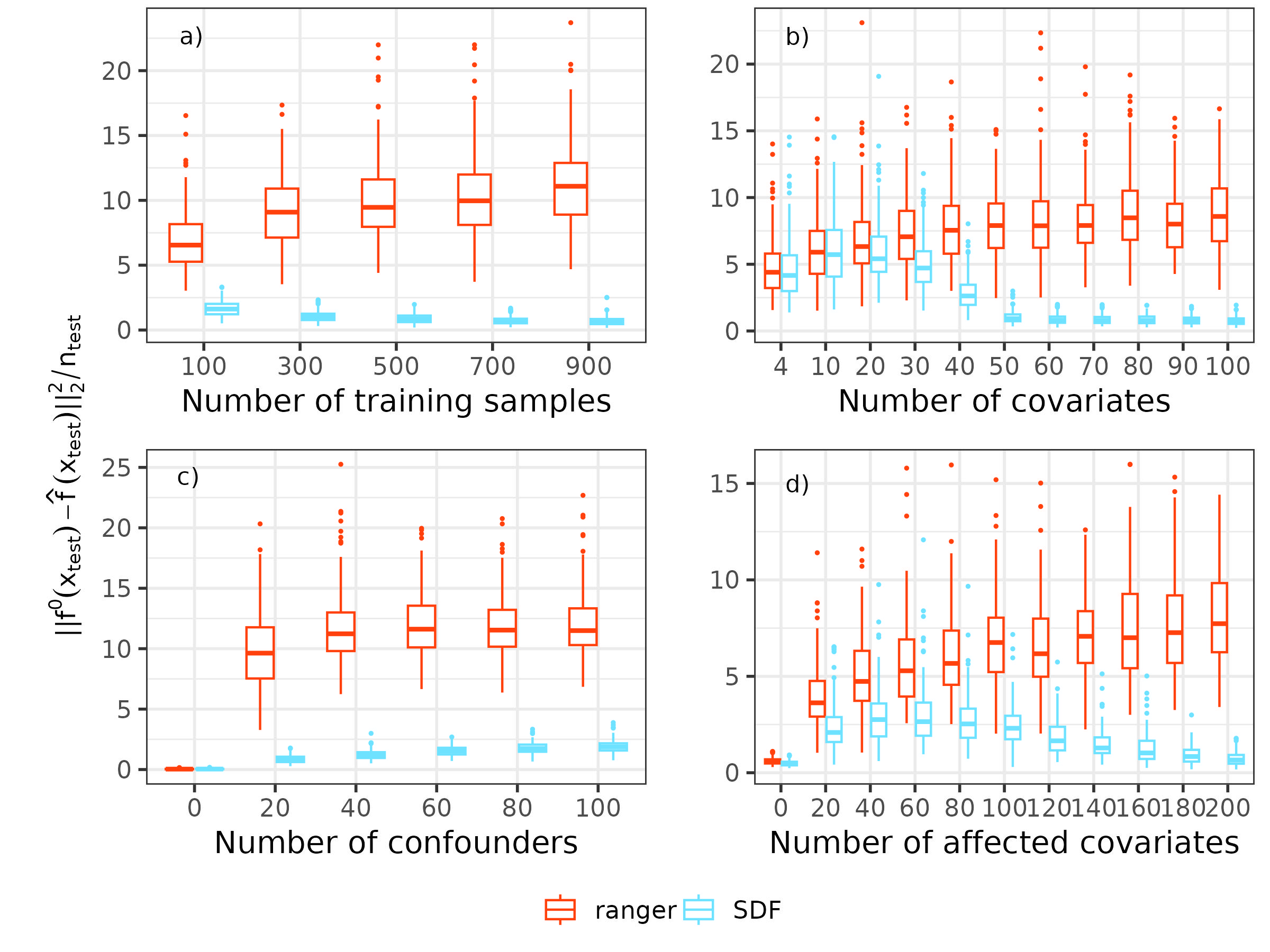}
  \caption{Mean squared error of the estimated causal function $\hat{f}(X)$ by classical Random Forests estimated by \emph{ranger} and the \emph{SDForests} depending on different simulation parameters. In subfigure a), we show how the performance depends on the sample size. In subfigure b), we show how the performance depends on the dimensionality of the observed data. The dependence of the performance on the amount of confounding is shown in subfigure c), where zero confounders corresponds to the classical setting without confounding. Subfigure d) shows the dependency of the performance on the number of affected covariates by the confounding, investigating the importance of dense confounding. Both algorithms estimate a hundred trees using $\mathsf{mtry} = \lfloor0.5p\rfloor$.}
  \label{fig:dims}
\end{figure}

The performance of the \emph{SDForests} and the classical Random Forests depending on different dimensions is shown in Figure \ref{fig:dims}. Theorem \ref{t:objective_prime} shows how the spectral objective \eqref{eq:spectral_objective} behaves as the number of variables $p$, and the number of observations $n$ grow to infinity. The term $R_n$ goes asymptotically with $1/\min(\sqrt{n}, \sqrt{p})$ to zero. So, we need both $n$ and $p$ to grow for consistency. In this simulation study, we examine the dependence of performance on these quantities in practice. Figure \ref{fig:dims}a shows the error distribution depending on the sample size. The \emph{SDForests} clearly outperforms the classical Random Forests in estimating the causal function, and the error decreases with a larger sample size. 

We see the dependence of the performance on the number of covariates in Figure \ref{fig:dims}b. At $p > 20$, we see how the \emph{SDForests} start to perform significantly better than the classical Random Forests. When increasing $p$ above 50, the error no longer decreases, suggesting that there is a certain threshold of $p$ beyond which deconfounding is successful. Even with only the four causal parents as covariates, the \emph{SDForests} do not lose performance in comparison to the classical Random Forests. 

With stronger confounding, estimating the causal function becomes increasingly difficult. Not only does the bias increase, but the variance in the response also increases. Therefore, we expect the error to increase with increasing confounding even for the \emph{SDForests}. Figure \ref{fig:dims}c shows this behavior with the simulation study, where we increase the number of confounding variables. For $q = 0$, the setting corresponds to the classical model without confounding, and both the classical Random Forests and the \emph{SDForests} perform equally well. It is important to note that we do not lose much, even when the data is not confounded, by applying the spectral deconfounding. However, we can gain a lot if there is confounding present. 

In Figure \ref{fig:dims}d, we follow up on the assumption of dense confounding. Here, instead of having random effects of the confounders on all the covariates, we simulate the data with only a random subsample of the covariates being affected by confounding. The denser the confounding becomes, the better the \emph{SDForests} perform. With around 200 covariates affected (40\%), the \emph{SDForests} start performing similarly as if there were no confounding, while with sparser confounding, \emph{SDForest} still outperform the classical Random Forests.

\section{Single-Cell Data}
\label{sec:gene}

We apply \emph{SDForest} and \emph{ranger} to compare the resulting model on the scRNA-seq single-cell gene expression dataset for the cell RPE1 generated by \cite{Replogle2022MappingPerturb-seq}. We use the preprocessed and filtered dataset provided by \cite{Chevalley2025AData}. As the response variable, we choose the gene EIF1 following the arguments of \cite{Shen2023Causality-orientedInterventions}, who conjecture that EIF1 might be a leaf node in the causal graph of genes. We use all other gene expressions ($p = 382$) as predictors. We consider observational data without any interventional gene knockouts ($n = 11485$ observations).

\begin{figure}[hbt!]
  \centering
  \includegraphics[width=0.5\textwidth]{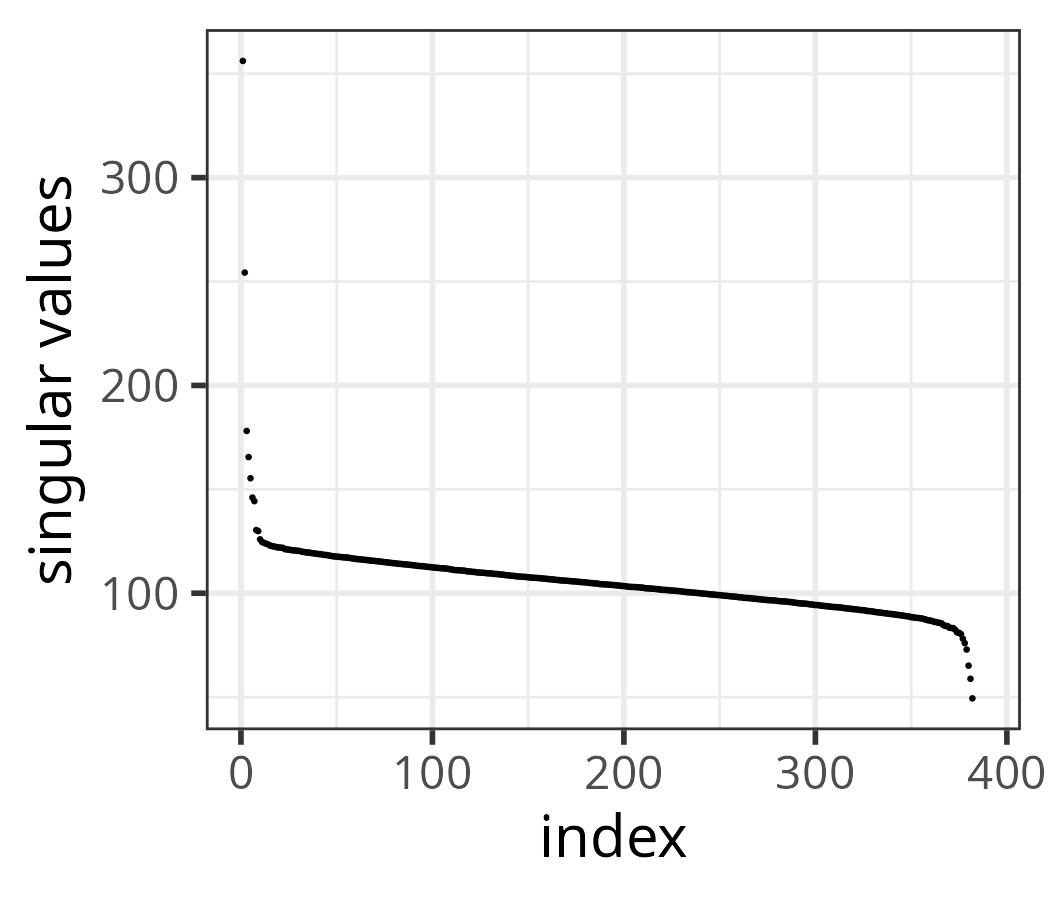}
  \caption{Singular values of the scaled predictors in the single-cell data.}
  \label{fig:cBench_sig}
\end{figure}

Examining the singular values of the scaled predictors, as shown in Figure \ref{fig:cBench_sig}, reveals a clear spike in the first few singular values. This is an indication of a factor structure that may induce dense confounding.

\begin{figure}[hbt!]
  \centering
  \includegraphics[width=1\textwidth]{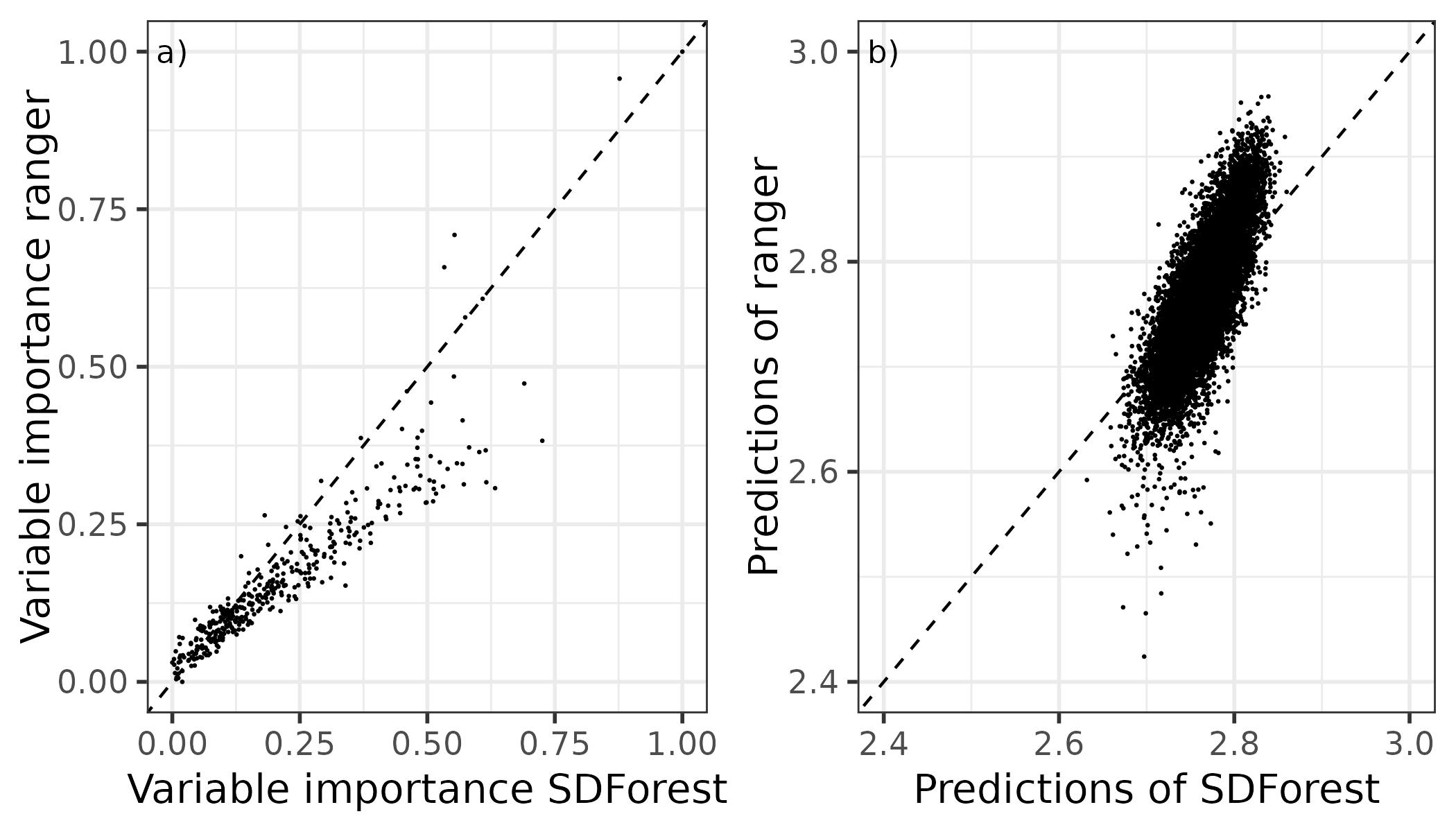}
    \caption{Comparison of the classical Random Forest estimated by \emph{ranger} and the \emph{SDForest} in the single-cell data. Subfigure a) shows the variable importance for both methods scaled to the interval $[0, 1]$. In subfigure b), the predictions $\hat{Y}$ of both methods are shown. The dashed 45-degree line corresponds to equality.}
  \label{fig:cBenchComparison}
\end{figure}

For both methods \emph{SDForest} and \emph{ranger}, we fit 500 trees and compare the estimated models in Figure \ref{fig:cBenchComparison}. Both models are fitted with $\mathsf{mtry} = 191$ and 1000 bootstrap samples per tree. The two methods agree on the variable importance for the most important genes. Thus, the most predictive genes, derived from \emph{ranger}, are not just arising due to dense confounding. When comparing the predictions (or predicted regression functions), there is a substantial correlation between the methods, but a slight shrinkage towards zero is observed with \emph{SDForest} due to optimizing the deconfounded loss, which shrinks the largest singular values. As a practical guideline, we advocate running and comparing \emph{SDForest} and \emph{ranger}, as it yields additional information and insights into whether variables and predictions are possibly induced by dense confounding or not. For this dataset,  we do not have strong evidence for dense confounding.

\begin{figure}[hbt!]
  \centering
  \includegraphics[width=1\textwidth]{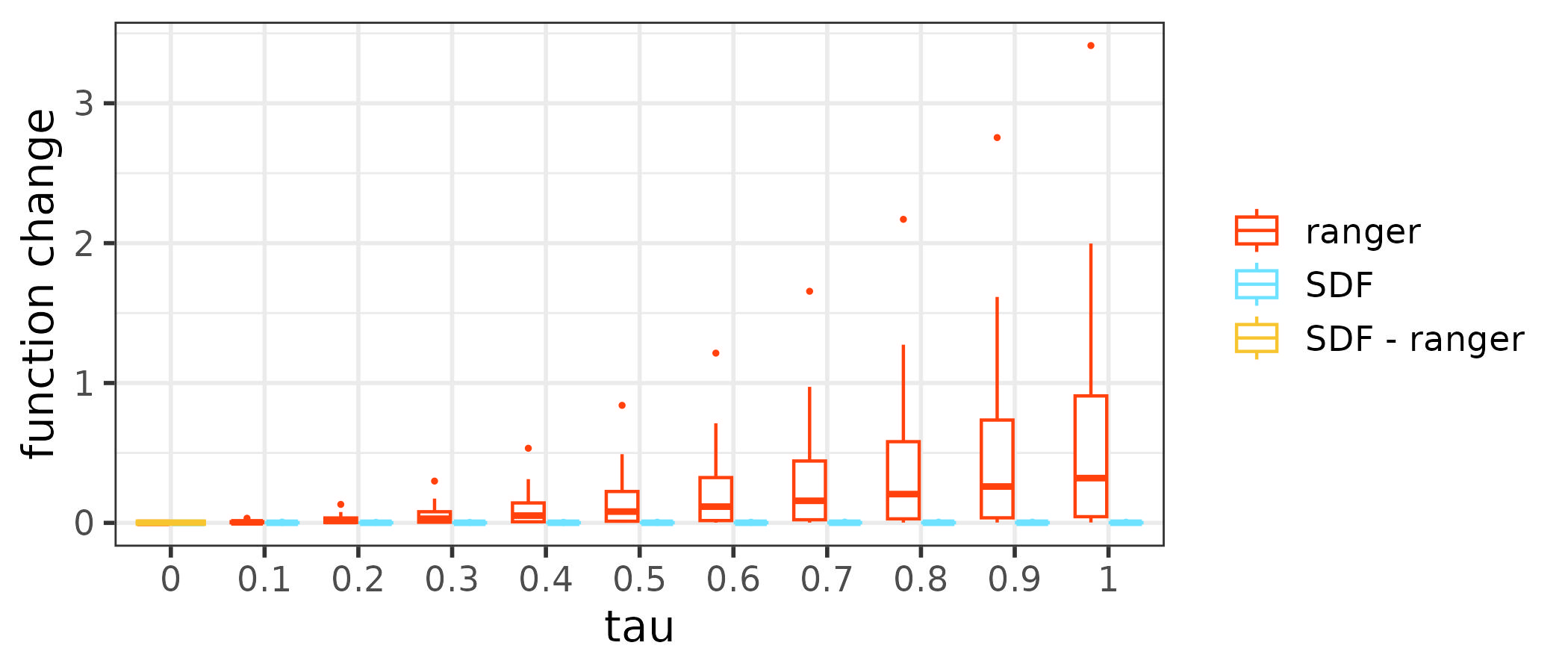}
  \caption{Change in prediction $\frac{\|f^{\text{method}}_{\tau}(\mathbf X_{\tau})-f^{\text{method}}_{0}(\mathbf X_{0}) \|_2^2}{n}$ of \emph{ranger} and \emph{SDForest} when perturbing the data with additional dense confounding. At $\tau$ = 0, we show the initial difference in prediction of \emph{ranger} and \emph{SDForest} when estimating and predicting on the unperturbed data $\frac{\|f^{\text{SDF}}_{0}(\mathbf X_{0})-f^{\text{\emph{ranger}}}_{0}(\mathbf X_{0}) \|_2^2}{n} = 0.0016$.}
  \label{fig:semiSim}
\end{figure}

To test for robustness against dense confounding, we synthetically perturb the original cleaned and filtered dataset. For this, we construct $\mathbf X_{\tau} \coloneqq \mathbf{X} + \mathbf{H} \Gamma \tau$ and $\mathbf Y_{\tau} \coloneqq Y + \mathbf{H} \delta \tau$, where $\mathbf{Y}$ is the original expression of EIF1 and $\mathbf{X}$ are the other original gene expressions used for prediction. The entries of $\mathbf{H} \in \mathbb{R}^{n}$, $\Gamma \in \mathbb{R}^{1 \times 382}$, and $\delta \in \mathbb{R}$ are sampled i.i.d. from $\mathcal N (0, 1)$ while we vary $\tau$ to increase the added dense confounding. We fit both methods to $\mathbf{X}_{\tau}$ and $\mathbf{Y}_{\tau}$ as $\tau$ increases and analyze how the estimated functions change. The change in the estimated functions is calculated as the change of the out-of-bag predictions of $f^{\text{\emph{ranger}}}_{\tau}(\mathbf X^{\tau})$ and $f^{\text{SDF}}_{\tau}(\mathbf X^{\tau})$ depending on $\tau$ $$\frac{\|f^{\text{method}}_{\tau}(\mathbf X_{\tau})-f^{\text{method}}_{0}(\mathbf X_{0}) \|_2^2}{n}.$$ We repeat the synthetic confounding of the data by sampling 20 times $\mathbf{H}$, $\Gamma$, and $\delta$ and increasing $\tau$. The distributions of the function changes are shown in Figure \ref{fig:semiSim}. With no added confounding, the difference between the two functions is small (see caption of Figure \ref{fig:semiSim}). This suggests that there is no major dense confounding present. As we increasingly add dense confounding, we clearly see how the estimated function by \emph{ranger} changes, while \emph{SDForest} demonstrates robustness against this added perturbation.

\section{Conclusion}
\label{sec:disc}

We propose the Spectrally Deconfounded Random Forest algorithm \emph{SDForest} with R-package \emph{SDModels} \if0\blind{\citep{Ulmer2025SDModels:Models}} \fi \if1\blind{blinded} \fi to estimate direct regression functions in high-dimensional data in the presence of dense hidden confounding. This can be used to screen for relevant covariates among a large set of variables, and the procedure provides robustness against unobserved confounding, gaining much in the confounded case but losing little if no confounding is present. \emph{SDModels} provides functions such as regularization paths and stability selection to screen for relevant covariates, as well as partial dependence plots to understand their relationship with the variable of interest. We demonstrate the empirical behavior of \emph{SDForests} in various settings, including challenging cases with sparse confounding, to illustrate some fundamental limitations (which are not found to be clearly worse than those of classical Random Forests). 

Many potentially confounded high-dimensional applications are 
about classification instead of regression. Currently, \emph{SDForest} is only applicable for regression tasks, and it is important to extend this methodology to classification. Another open question is whether one can combine spectral deconfounding with Quantile Regression Forests \citep{Meinshausen2006QuantileForests} or Distributional Random Forests \citep{Hothorn2021PredictiveForests, Cevid2022DistributionalRegression} to gain access to prediction intervals or construct confidence intervals using techniques as in 
\cite{Guo2022DoublyConfounding} for the linear case and \cite{Naf2023ConfidenceForests} using Random Forests.

\section*{Supplementary Materials}

\textbf{Appendix A:} Approximation of splitting criteria\\
\textbf{Appendix B:} Visualization of Spectral Transformation of singular values\\
\textbf{Appendix C:} Notes on Non-linear Confounding\\
\textbf{Appendix D:} Proofs of all theoretical results\\
\textbf{Code:} All the code used for this paper is available here: \if0\blind{\url{https://github.com/markusul/SDForest-Paper}} \fi \if1\blind{blinded} \fi \\
\textbf{R-package for \emph{SDForests}:} R-package \if0\blind{SDModels} \fi \if1\blind{\emph{blinded}} \fi provides software for non-linear spectrally deconfounded models: \if0\blind{\url{https://github.com/markusul/SDModels}} \fi \if1\blind{blinded} \fi

\section*{Acknowledgment}

\if0\blind{We thank Mathieu Chevalley for his support in applying our method to the single-cell dataset. We would also like to thank Andrea Nava, Malte Londschien, Minhui Jiang, and Christoph Schultheiss for their helpful discussions and comments.} \fi \if1\blind{blinded} \fi

\section*{Funding}

\if0\blind{M. Ulmer and C. Scheidegger gratefully acknowledge financial support from the Swiss National Science Foundation, grant no. 214865.} \fi \if1\blind{blinded} \fi

\section*{Disclosure statement}

The authors report there are no competing interests to declare

\newpage

\begin{center}

{\large\bf SUPPLEMENTARY MATERIAL}

\end{center}

\appendix
\setcounter{theorem}{0}

\section{Approximation of splitting criteria}

In the $M$th iteration of Algorithm 1, we need to find, in each region, the split that reduces the loss the most. This means we have $ M$ regions in each iteration, each with an optimal split to choose from. When we split the region at the optimal point, we obtain two new regions. Now, due to the spectral transformation, the samples, as well as the different regions containing a subset of the samples, are not independent. Therefore, to truly find the next best split, we would need to determine the optimal split and its corresponding loss decrease in each region anew.

\begin{figure}[H]
  \centering
  \includegraphics[width=0.5\textwidth]{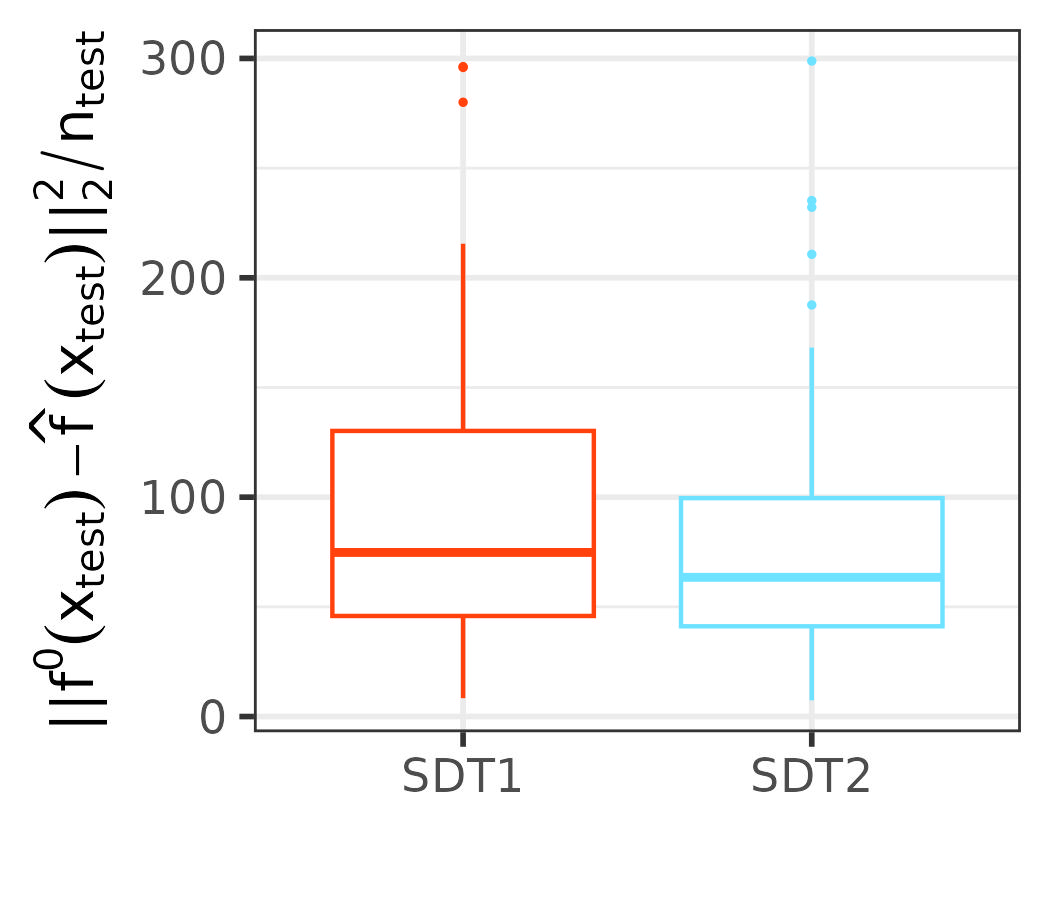}
  \caption{Mean squared error of the estimated causal function $\hat{f}(X)$ by SDTree using $\mathsf{cp} = 0.01$. SDT1 corresponds to Algorithm 1 while SDT2 uses $\mathcal{B} \gets (1, \ldots, M + 1)$ in line 22 instead of only the new partitions. The data is simulated using the confounding model (1) and with the same parameter as in Section 5 but using a random regression tree for $f^0$ as in Equation 8.}
  \label{fig:fast}
\end{figure}

This is done in Figure \ref{fig:fast} using the method \emph{SDT2}. In our studies and as the default in the R-package, we only estimate the optimal split and its loss decrease for the two new regions and reuse all the other estimates from the previous iteration. This saves substantial computational time, and we argue that a previously good split stays reasonable. We compare the performance of the two options in Figure \ref{fig:fast}. We simulate data following the confounding model (1) and using the same parameters as in Section 5, but with a random regression tree for $f^0$ as in Equation 8. The random regression tree is grown using random splits until there are ten leaf nodes. While the reestimation of all splits in \emph{SDT2} might be a bit better, we do not see a significant decrease in performance when using the computationally much more efficient approximation \emph{SDT1}. 

In the case of smooth underlying regression functions, we expect to see similar behavior (because we can consider the approximation error of a smooth function with a tree that remains unaffected and the estimation error of the best-approximated tree function, which is analogous to the discussion above).

\section{Transformation of singular values}

\begin{figure}[H]
  \centering
  \includegraphics[width=1\textwidth]{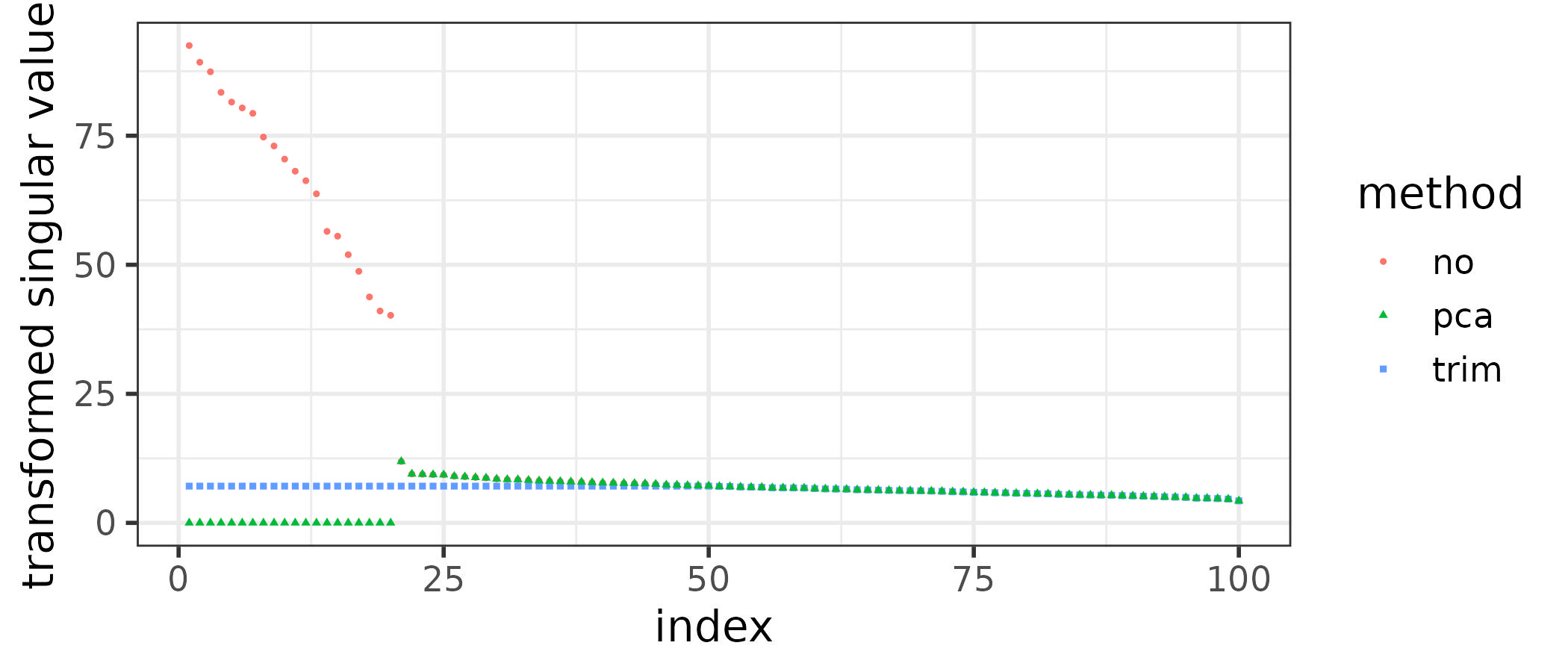}
  \caption{Singular values of $\mathbf X$, $Q_{trim} \mathbf X$, and $Q_{pca} \mathbf X$. The data is simulated using the confounding model (1) with the same parameters as in Section 5 ($q = 20$ hidden factors), but with $p = 100$ to increase visibility.}
  \label{fig:sig}
\end{figure}

The spectral transformation described in Section 3 shrinks the first few singular values to decrease confounding bias. This is visualized in Figure \ref{fig:sig} for the PCA adjustment and the trim-transform. PCA adjustment removes all the signal from the first 20 principal components (we assume here that the number of 20 hidden factors is known). This should not reduce the signal of $f^0(X)$ as it lies in the span of a sparse set of covariates. However, choosing the correct number of principal components in real data is subtle \citep{Owen2016Bi-Cross-ValidationAnalysis} and may result in unwanted removal of signal of $f^0(X)$. 

The trim-transform is much less sensitive to the problem with the unknown number of confounders, as it only limits the top half of the singular values to their median and does not entirely remove any principal components and their associated signals.

\section{Non-linear Confounding}

In Equation 1 and Section 3.1, we assumed that the hidden confounder affects both the $X$ and $Y$ linearly. This assumption is not testable and might not hold in practice. We present here some heuristic arguments for when and why spectral deconfounding could still work well with nonlinear confounding. Without loss of generality, assume that $H$ is univariate and assume that the data is generated according to
\begin{equation} \label{eq:nlconfounding}
    Y = f(X) + d(H) + \nu, \quad X_j = g_j(H) + E_j,\, j = 1,\ldots, p,
\end{equation}    
for some (potentially) nonlinear functions $d$ and $g_j$, $j=1,\ldots, p$ from $\mathbb R\to\mathbb R$. Assume that $g_j(\cdot), j = 1,\ldots, p$ and $d(\cdot)$ can be well-approximated by a common set of basis functions, $(b_k(\cdot))_{k=1}^K$, i.e., $g_j(\cdot) \approx \sum_{k = 1}^K \Gamma_{k, j} b_k(\cdot)$, $j=1,\ldots, p$ and $d(\cdot) \approx \sum_{k=1}^K \delta _k b_k(\cdot)$. Let $B = (b_1(H), \ldots, b_K(H))^T\in \mathbb R^K$. Then, it approximately holds that $$Y \approx f(X) + \delta^T B + \nu, \quad X \approx \Gamma^T B + E,$$ i.e., we are approximately in the setting of Equation 1 with linear confounding with $H$ being replaced by $B$ (and without loss of generality, by orthogonalization, we can assume $\mathrm{Cov}(B) = I$). 

If $\Gamma$ is dense, it is reasonable that spectral deconfounding still works well. Intuitively, $\Gamma$ will be dense if the $g_j$ are all ``sufficiently different''. Moreover, $d(\cdot)$ should be ``similar'' to the $g_j$ such that there is an approximation with a common basis. The number $K$ in the basis approximations is then analogous to the number of confounding variables. If we assume that the functions $d(\cdot)$ and $g_j(\cdot)$, $j=1,\ldots, p$ are not too complicated, it is still reasonable to assume that $K$ is small and we do not need to know it.

Empirically, we simulate data similar to Section 5. In addition to the non-linear function $f^0(X)$, we simulate non-linear confounding using Equation \ref{eq:nlconfounding} where we simulate $g_j$ and $d$ using different random functions using the Fourier basis (also using Equation 13). For the confounding effect on $X$ and $Y$, we use $K = 12$ basis functions and sample all the coefficients uniformly on $[-1, 1]$ for the effect on $X$ and on $[-2, 2]$ for the effect on $Y$. 

Here, we use $q = 1$, $p = 300$, $n = 500$, and let only one covariate affect the response $Y$. For reasonable noise level and confounding strength, we sample $\delta \in \mathbb R$ i.i.d.\ from a Gaussian with mean zero and $\sigma = 2$ and the additional noise $\nu \in \mathbb R ^ n$ from a Gaussian with mean zero and $\sigma_{\nu} = 0.01$. All the other parameters in the simulation stay the same as in Section 5. 

\begin{figure}[H]
  \centering
  \includegraphics[width=1\textwidth]{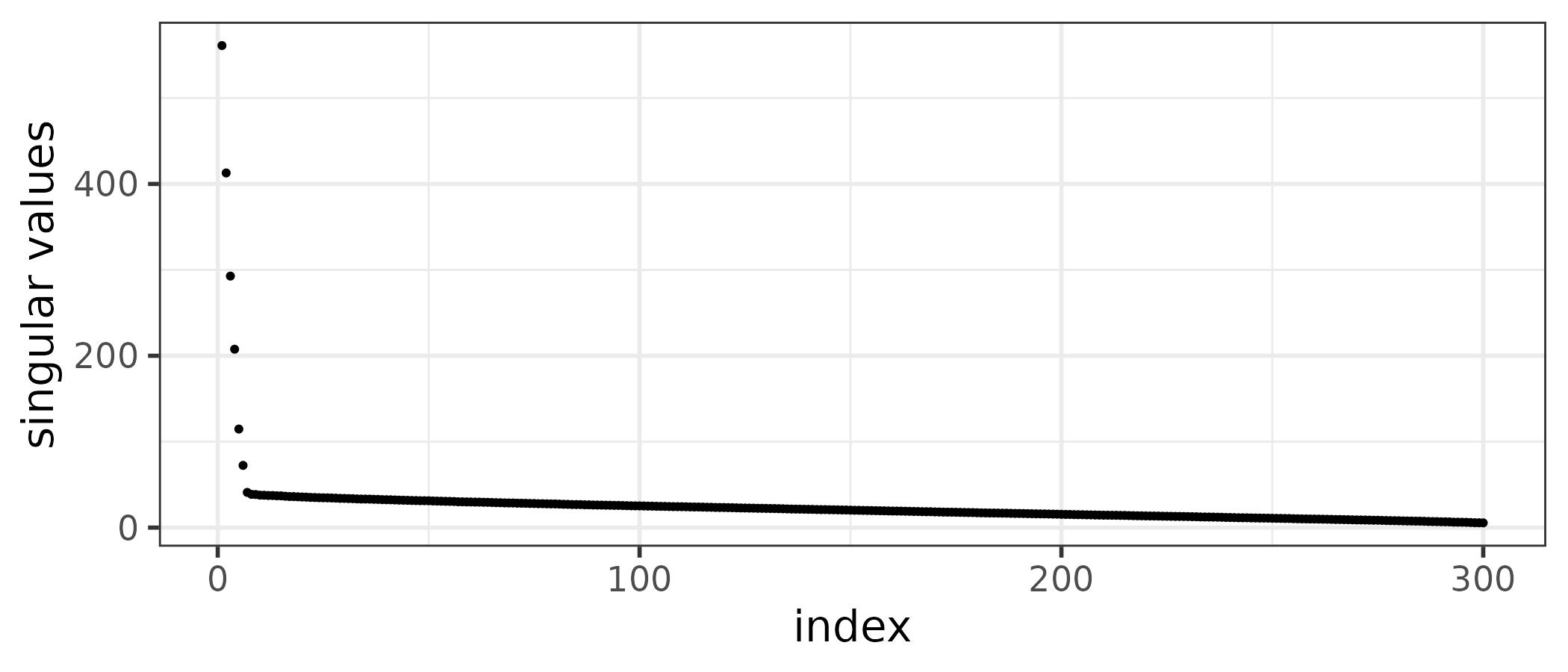}
  \caption{Singular values of a random realization of $\mathbf X$ affected non-linearly by a single confounder.}
  \label{fig:sig_nl2}
\end{figure}

In Figure \ref{fig:sig_nl2}, we show the singular values of $\mathbf{X}$ of a random realization. We observe that six instead of just one singular value spike due to the non-linear confounding. Apparently, this number is lower than the number of basis functions $K = 12$, and the spiking effect is only visible for the first six components.

\begin{figure}[hbt!]
  \centering
  \includegraphics[width=1\textwidth]{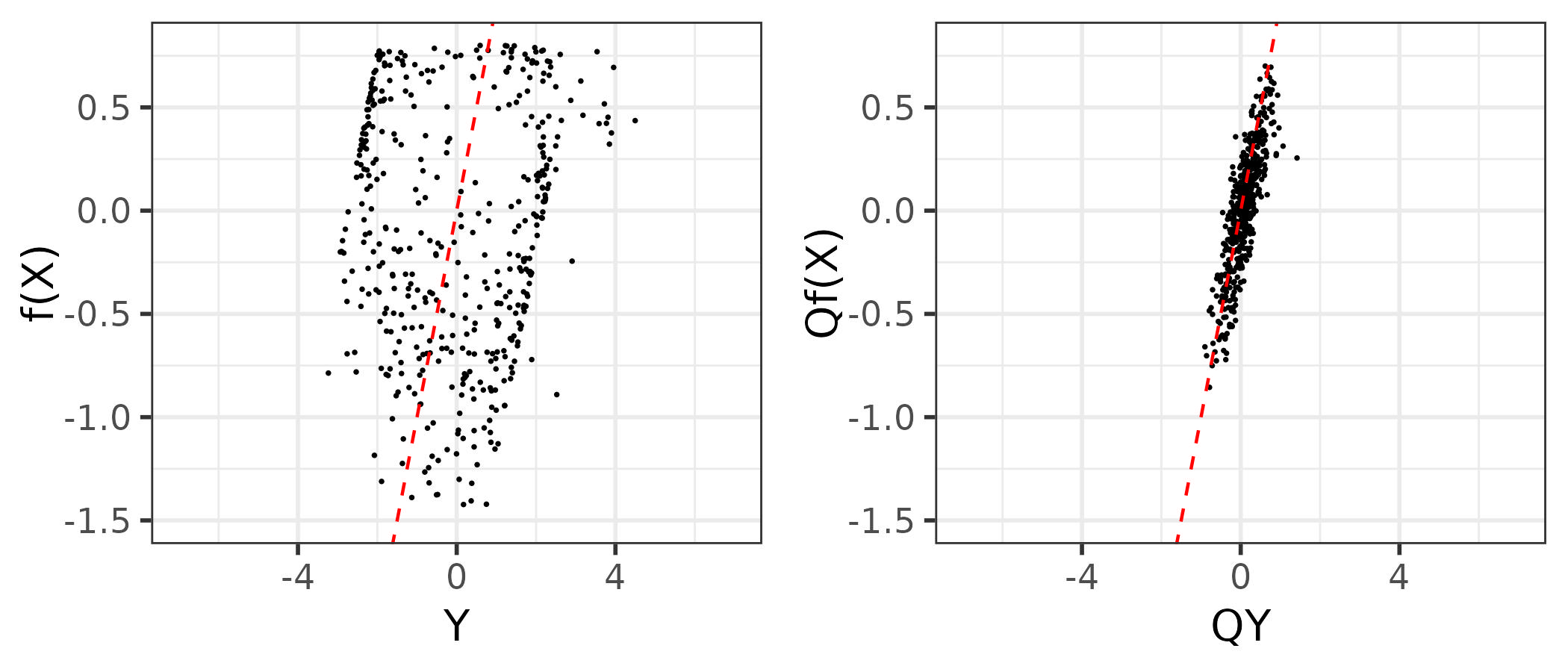}
  \caption{A random realization of the non-linearly confounded process. On the left, we show $f^0(\mathbf{X})$ against $\mathbf{Y}$; on the right, the spectrally transformed versions are shown against each other, that is, $Qf^0(\mathbf{X})$ versus $Q\mathbf{Y}$. In both visualizations, the line with a slope equal to one, which corresponds to perfect correlation, is shown as a dashed line. This is the same visualization as in Figure 2 for the linear confounding.}
  \label{fig:pt_nl2}
\end{figure}

\begin{figure}[H]
  \centering
  \includegraphics[width=1\textwidth]{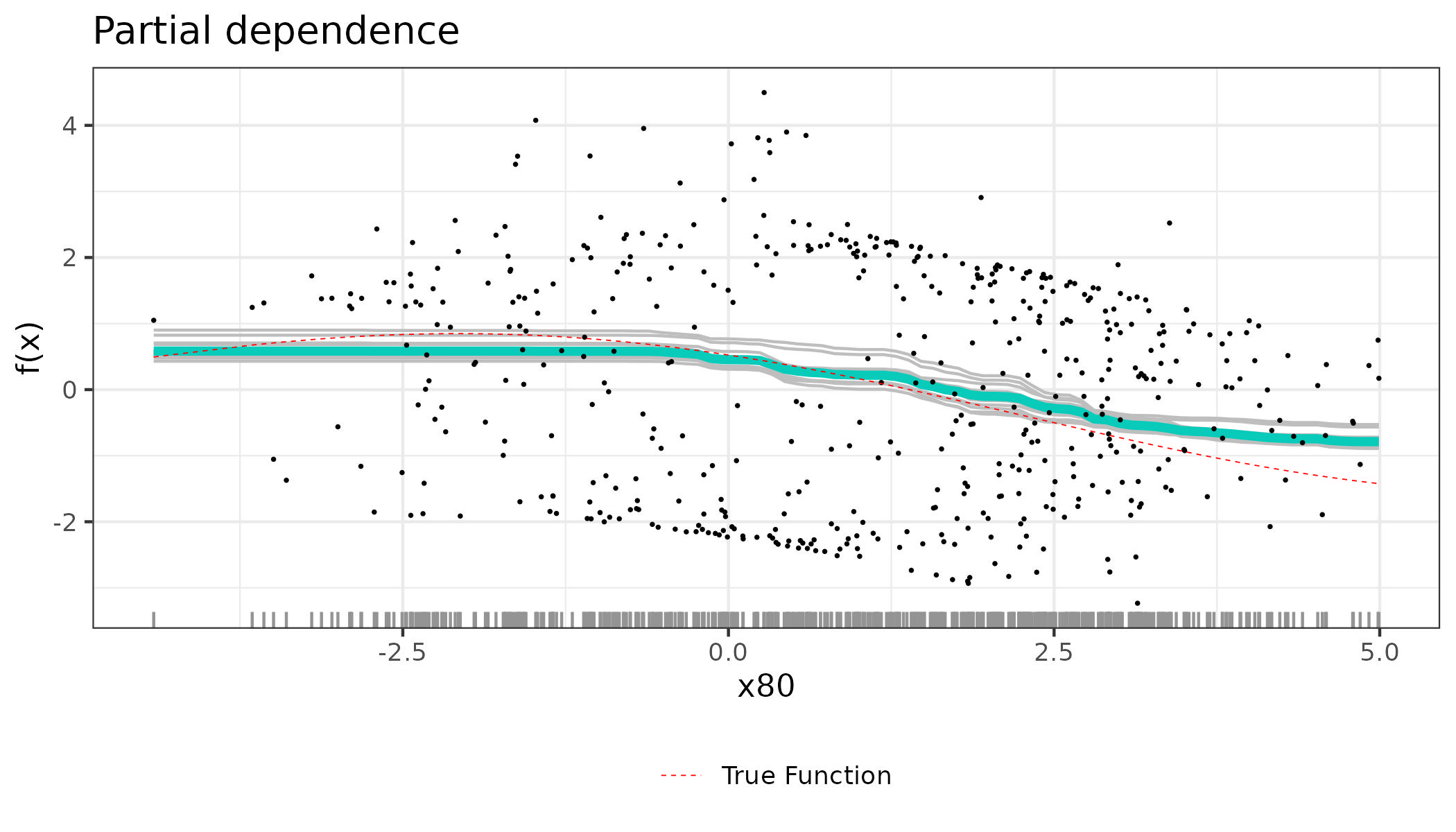}
  \caption{Partial dependence plots of the estimated \emph{SDForest} for the true causal parent of the response $Y$ in the non-linearly confounded setting. The dashed line is the corresponding true partial causal function. The light lines show the observed empirical partial functions for 19 randomly selected observations, and the thick line is the average of all observed partial functions.}
  \label{fig:dep_nl2}
\end{figure}

In this setting, $\mathbf{Y}$ is an even worse approximation for $f^0(\mathbf{X})$ compared to the linear confounding setting. However, the spectral transformation still yields a clear correlation between $Q\mathbf{Y}$ and $Qf^0(\mathbf{X})$, as shown in Figure \ref{fig:pt_nl2}. In Figure \ref{fig:dep_nl2}, we show the dependence of $\mathbf{Y}$, $f^0(\mathbf{X})$, and $f^{SDForest}(\mathbf{X})$ on the single causal parent $X_{80}$. The observations (points in the figure) show a complicated dependency on $X_{80}$. 

\begin{figure}[H]
  \centering
  \includegraphics[width=0.5\textwidth]{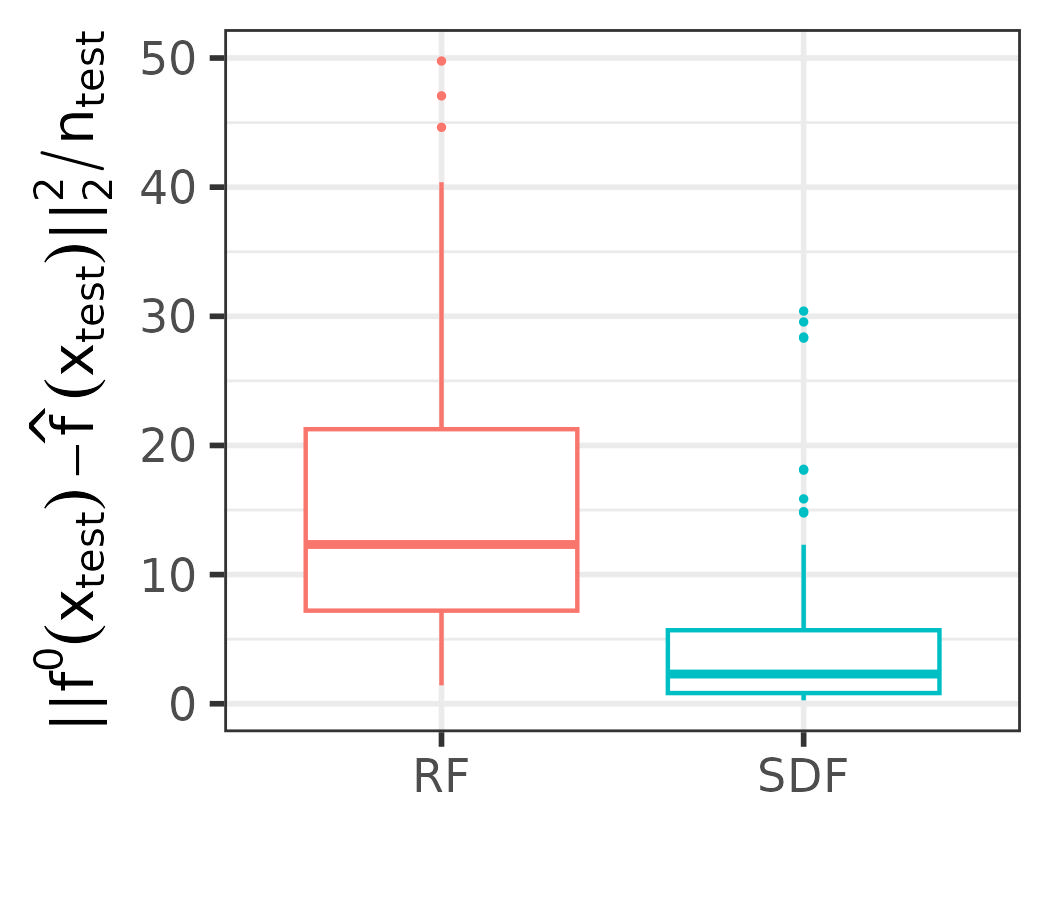}
  \caption{Mean squared error of the estimated causal function $\hat{f}(X)$ by classical Random Forests estimated by \emph{ranger} and the \emph{SDForests} in the non-linearly confounded setting.}
  \label{fig:perf_nl2}
\end{figure}

At the same time, the estimated relationship by \emph{SDForest}, represented by the thick line, stays close to the actual causal function (dashed line). We repeat this simulation a hundred times and report the test performance, using 500 data points, of classical Random Forests and the \emph{SDForest} in estimating the actual function $f^0(\mathbf{X})$. The distribution of these performances is shown in Figure \ref{fig:perf_nl2}, where we clearly see that \emph{SDForest} outperforms the classical Random Forests in this specific non-linear confounding setting as well.

\section{Proofs}

\begin{theorem} \label{t:objective_prime}
    Assume the confounding model (1) and assume that the conditions (4), (5), (6) and (7) hold. Then, it holds that
    
    \begin{equation*}
        \frac{\|Q(\mathbf{Y} - f(\mathbf{X}))\|_2}{\sqrt n} = \frac{\|Q(f^0(\mathbf{X}) - f(\mathbf{X}) + \nu)\|_2}{\sqrt{n}} + R_n
    \end{equation*}

    where $R_n = \mathcal{O}_{\mathbb P}\left( \frac{\|\delta\|_2}{\min(\sqrt{n}, \sqrt{p})}\right)$.
\begin{proof}
As in \cite{Guo2022DoublyConfounding} and \cite{Cevid2020SpectralModels}, let $b=\arg\min_{b'\in \mathbb R^p}\mathbb E[(X^T b' - H^T\delta)^2] = \mathbb E[XX^T]^{-1}\mathbb E[X H^T\delta]$, i.e. $X^T b$ is the $L_2$ projection of $H^T\delta$ onto $X$. By (1) and the triangle inequality, we have that
        \begin{equation*}
        \begin{split}
            &\left|\frac{\|Q(\mathbf{Y} - f(\mathbf{X}))\|_2}{\sqrt n} - \frac{\|Q(f^0(\mathbf{X}) - f(\mathbf{X}) + \nu)\|_2}{\sqrt n}\right|\\
            & \leq \frac{\|Q(\mathbf{Y} - f(\mathbf{X})) - (Q(f^0(\mathbf{X}) - f(\mathbf{X}) + \nu))\|_2}{\sqrt n}
            \\
            & = \frac{\|Q(f^0(\mathbf{X}) + \mathbf{H}\delta + \nu - f(\mathbf{X}))- (Q(f^0(\mathbf{X}) - f(\mathbf{X}) + \nu))\|_2}{\sqrt n} \\
            & = \frac{\|Q(f^0(\mathbf{X}) - f(\mathbf{X}) + \mathbf{X}b + (\mathbf{H}\delta - \mathbf{X}b) + \nu)- (Q(f^0(\mathbf{X}) - f(\mathbf{X}) + \nu))\|_2}{\sqrt n} \\
            &\leq \frac{\|Q\mathbf X b\|_2}{\sqrt n} + \frac{\|Q(\mathbf H\delta -\mathbf Xb)\|_2}{\sqrt n}
        \end{split}
        \end{equation*}

        From Lemma \ref{l:b_prime} below, we have that $\|b\|_2=\mathcal O(\|\delta\|_2/\sqrt p)$ and hence using (7)

        \begin{equation*}
            \frac{\|Q\mathbf{X}b\|_2}{\sqrt n} \leq \frac{1}{\sqrt n}\lambda_{\max}(Q\mathbf{X}) \|b\|_2 = \mathcal{O}\left(\|\delta\|_2\frac{\max(\sqrt n, \sqrt p)}{\sqrt{np}}\right) = \mathcal O\left(\frac{\|\delta\|_2}{\min(\sqrt n, \sqrt p)}\right).
        \end{equation*}
             By Lemma \ref{l:bias_prime} below, the second term behaves as
        \begin{equation*}
            \frac{1}{\sqrt n} \|Q(\mathbf{H}\delta - \mathbf{X}b)\|_2 = \mathcal{O}_\mathbb{P}\left(\frac{\|\delta\|_2}{\sqrt p}\right),
        \end{equation*}
        which concludes the proof.
    \end{proof}
\end{theorem}

\begin{lemma}[Parts of Lemma 6 in \cite{Cevid2020SpectralModels}] \label{l:b_prime}
 Assume that the confounding model (1) satisfies the assumptions (4), (5) and (6). Then we have,

\begin{equation*}
    \|b\|_2^2 = \|\mathbb E[XX^T]^{-1}\mathbb E[XH^T]\delta\|_2^2 \leq \textup{cond}(\Sigma_E) \cdot \frac{\|\delta\|_2^2}{\lambda_{\min}(\Gamma)^2} = \mathcal{O}\left(\frac{\|\delta\|_2^2}{p}\right).
\end{equation*}
\end{lemma}

\begin{lemma} \label{l:bias_prime}
Under the conditions of Theorem \ref{t:objective_prime},

\begin{equation*}
    \frac{1}{n} \|Q(\mathbf{H}\delta - \mathbf{X}b)\|_2^2 = \mathcal{O}_\mathbb{P}\left(\frac{1}{p}\right)
\end{equation*}

\begin{proof}
    Observe that $\frac{1}{n}\|Q(\mathbf{H}\delta - \mathbf{X}b)\|_2^2 \leq \frac{1}{n}\|Q\|_{op}^2\|\mathbf{H}\delta - \mathbf{X}b\|_2^2$, so it suffices to show that $\frac{1}{n}\|\mathbf{H}\delta - \mathbf{X}b\|_2^2 = \mathcal{O}_\mathbb{P}\left(\frac{1}{p}\right)$. By Markov's inequality, it suffices to show

    \begin{equation}
        \mathbb{E} \left[\frac{1}{n}\|\mathbf{H}\delta - \mathbf{X}b\|_2^2 \right] = \mathcal{O}\left(\frac{1}{p}\right).
    \end{equation}

    For this, we follow the arguments of \cite{Guo2022DoublyConfounding}. Our term $\mathbb{E} \left[\frac{1}{n}\|\mathbf{H}\delta - \mathbf{X}b\|_2^2 \right]$ corresponds to $\Delta$ in (47) in A.4. there. We can follow the proof of (35) in Lemma 2 given in Section C.4 in \cite{Guo2022DoublyConfounding}. For this, we write

    \begin{equation*}
        \frac{1}{n}\|\mathbf{H}\delta - \mathbf{X}b\|_2^2 = \frac{1}{n} \sum_{i=1}^n (\mathbf{H}_i^T \delta - \mathbf{X}_i^T b)^2
    \end{equation*}
    Hence, 
    \begin{equation*}
    \begin{split}
        \mathbb{E} \left[\frac{1}{n}\|\mathbf{H}\delta - \mathbf{X}b\|_2^2 \right] & = \frac{1}{n} \sum_{i=1}^n \mathbb{E}[(\mathbf{H}_i^T \delta - \mathbf{X}_i^T b)^2] \\
        & = \mathbb{E} [(H^T\delta - X^Tb)^2] \\
        & = \mathbb{E}[\delta^T HH^T \delta - 2 \delta^T HX^Tb + b^TXX^Tb] \\
        & = \delta^T \mathbb{E}[HH^T]\delta - 2 \delta^T \mathbb{E} [HX^T]b + b^T \mathbb{E}[XX^T]b \\
        & = \delta^T \mathbb{E}[HH^T]\delta - 2 \delta^T \mathbb{E} [HX^T] \mathbb{E}[XX^T]^{-1}\mathbb{E}[XH^T]\delta + \delta^T \mathbb{E}[HX^T]\mathbb{E}[XX^T]^{-1}\mathbb{E}[XH^T]\delta \\
        & = \delta^T \mathbb{E}[HH^T]\delta - \delta^T \mathbb{E} [HX^T] \mathbb{E}[XX^T]^{-1}\mathbb{E}[XH^T]\delta \\
        & = \delta^T (I_{q} - \Gamma(\Gamma^T \Gamma + \Sigma_E)^{-1} \Gamma^T) \delta
    \end{split}
    \end{equation*}
    where we used the definition of $b$ and (4). As in equation (134) in the supplementary materials in \cite{Guo2022DoublyConfounding}, using Woodbury's identity, we have

    \begin{equation*}
        I_{q} - \Gamma(\Gamma^T \Gamma + \Sigma_E)^{-1} \Gamma^T = (I_q + \Gamma\Sigma_E^{-1}\Gamma^T)^{-1}.
    \end{equation*}
    Let $C = \Sigma_E^{-1/2}\Gamma^T$ and $C = U_C D_C V_C^T$ be the singular values decomposition of $C$. Then, we have

    \begin{equation*}
        \begin{split}
            \mathbb{E} \left[\frac{1}{n}\|\mathbf{H}\delta - \mathbf{X}b\|_2^2 \right] & = \delta^T (I_q + \Gamma\Sigma_E^{-1}\Gamma^T)^{-1} \delta \\
            & \leq \|\delta\|_2^2 \lambda_{\max}\left((I_q + \Gamma^T\Sigma_E^{-1}\Gamma)^{-1}\right) \\
            & = \|\delta\|_2^2 \lambda_{\max}\left((I_q + C^TC)^{-1}\right) \\
            & = \|\delta\|_2^2 \lambda_{\max}\left(V_C(I_q + D_C^TD_C)^{-1}V_C^T\right) \\
            & = \|\delta\|_2^2 (1 + \lambda_{\min}(D_C^TD_C))^{-1} \\
            & = \|\delta\|_2^2 (1 + \lambda_{\min}(C^TC))^{-1} \\
            & \leq \|\delta\|_2^2\lambda_{\min}(C^TC)^{-1}
        \end{split}
    \end{equation*}

    Note that

    \begin{equation*}
        \begin{split}
            \lambda_{\min}(C^TC) & = \min_{x \neq 0} \frac{x^TC^TCx}{x^Tx} \\
            & = \min_{x \neq 0} \frac{x^T\Gamma \Sigma_E^{-1}\Gamma^T x}{x^T\Gamma\Gamma^T x} \frac{x^T\Gamma\Gamma^T x}{x^Tx} \\
            & \geq \lambda_{\min}(\Sigma_E^{-1})\lambda_{\min}(\Gamma\Gamma^T) \\
            & = \lambda_{\max}(\Sigma_E)^{-1}\lambda_{\min}(\Gamma)^2
        \end{split}
    \end{equation*}

    Hence, 

    \begin{equation*}
        \mathbb{E} \left[\frac{1}{n}\|\mathbf{H}\delta - \mathbf{X}b\|_2^2 \right] \leq \|\delta\|_2^2 \lambda_{\max}(\Sigma_E) \lambda_{\min}(\Gamma)^{-2} = \mathcal O \left(\frac{\|\delta\|_2^2}{p}\right)
    \end{equation*}   
\end{proof}
\end{lemma}

\begin{description}

\item[Code:] All the code used for this paper is available here: \if0\blind{\url{https://github.com/markusul/SDForest-Paper}} \fi \if1\blind{blinded} \fi 

\item[R-package for \emph{SDForests}:] R-package \if0\blind{SDModels} \fi \if1\blind{\emph{blinded}} \fi provides software for non-linear spectrally deconfounded models: \if0\blind{\url{https://github.com/markusul/SDModels}} \fi \if1\blind{blinded} \fi

\end{description}

\bibliography{references}

@article{Chevalley2025AData,
    title = {{A large-Scale Benchmark for Network Inference from Single-Cell Perturbation Data}},
    year = {2025},
    journal = {Communications Biology},
    author = {Chevalley, Mathieu and Roohani, Yusuf and Mehrjou, Arash and Leskovec, Jure and Schwab, Patrick},
    pages = {2399--3642},
    volume = {8},
    arxivId = {2210.17283}
}

@article{Fan2024AreAdequate,
    title = {{Are Latent Factor Regression and Sparse Regression Adequate?}},
    year = {2024},
    journal = {Journal of the American Statistical Association},
    author = {Fan, Jianqing and Lou, Zhipeng and Yu, Mengxin},
    number = {546},
    pages = {1076--1088},
    volume = {119},
    doi = {10.1080/01621459.2023.2169700},
    issn = {0162-1459}
}

@article{Breiman1996BaggingPredictors,
    title = {{Bagging Predictors}},
    year = {1996},
    journal = {Machine Learning},
    author = {Breiman, Leo},
    number = {2},
    pages = {123--140},
    volume = {24},
    doi = {10.1007/BF00058655},
    issn = {0885-6125}
}

@article{Owen2016Bi-Cross-ValidationAnalysis,
    title = {{Bi-Cross-Validation for Factor Analysis}},
    year = {2016},
    journal = {Statistical Science},
    author = {Owen, Art B. and Wang, Jingshu},
    number = {},
    pages = {119--139},
    volume = {31},
    doi = {10.1214/15-STS539},
    issn = {0883-4237}
}

@article{Leek2007CapturingAnalysis,
    title = {{Capturing Heterogeneity in Gene Expression Studies by Surrogate Variable Analysis}},
    year = {2007},
    journal = {PLoS Genetics},
    author = {Leek, Jeffrey T and Storey, John D},
    number = {9},
    pages = {1--12},
    volume = {3},
    doi = {10.1371/journal.pgen.0030161},
    issn = {1553-7404}
}

@article{Peters2016CausalIntervals,
    title = {{Causal Inference by Using Invariant Prediction: Identification and Confidence Intervals}},
    year = {2016},
    journal = {Journal of the Royal Statistical Society. Series B (Statistical Methodology)},
    author = {Peters, Jonas and B{\"{u}}hlmann, Peter and Meinshausen, Nicolai},
    number = {5},
    pages = {947--1012},
    volume = {78},
    publisher = {[Royal Statistical Society, Wiley]},
    issn = {13697412, 14679868}
}

@article{Shen2023Causality-OrientedInterventions,
    title = {{Causality-Oriented Robustness: Exploiting General Additive Interventions}},
    year = {2023},
    journal = {arXiv:2307.10299},
    author = {Shen, Xinwei and B{\"{u}}hlmann, Peter and Taeb, Armeen},
    arxivId = {2307.10299}
}

@book{Pearl2009Causality:Inference,
    title = {{Causality: Models, Reasoning, and Inference}},
    year = {2009},
    author = {Pearl, Judea},
    edition = {2},
    number = {},
    volume = {},
    publisher = {Cambridge University Press},
    address = {Cambridge, United Kingdom}
}

@book{Breiman2017ClassificationTrees,
    title = {{Classification and Regression Trees}},
    year = {2017},
    author = {Breiman, Leo and Friedman, Jerome H. and Olshen, Richard A. and Stone, Charles J.},
    month = {10},
    publisher = {Chapman and Hall/CRC},
    address = {New York},
    isbn = {9781315139470},
    doi = {10.1201/9781315139470}
}

@article{Naf2023ConfidenceForests,
    title = {{Confidence and Uncertainty Assessment for Distributional Random Forests}},
    year = {2023},
    journal = {Journal of Machine Learning Research},
    author = {N{\"{a}}f, Jeffrey and Emmenegger, Corinne and B{\"{u}}hlmann, Peter and Meinshausen, Nicolai},
    number = {366},
    pages = {1--77},
    volume = {24}
}

@article{Cevid2022DistributionalRegression,
    title = {{Distributional Random Forests: Heterogeneity Adjustment and Multivariate Distributional Regression}},
    year = {2022},
    journal = {Journal of Machine Learning Research},
    author = {{\'{C}}evid, Domagoj and Michel, Loris and N{\"{a}}f, Jeffrey and B{\"{u}}hlmann, Peter and Meinshausen, Nicolai},
    number = {333},
    pages = {1--79},
    volume = {23}
}

@article{Guo2022DoublyConfounding,
    title = {{Doubly Debiased Lasso: High-Dimensional Inference Under Hidden Confounding}},
    year = {2022},
    journal = {The Annals of Statistics},
    author = {Guo, Zijian and {\'{C}}evid, Domagoj and B{\"{u}}hlmann, Peter},
    number = {},
    pages = {1320--1347},
    volume = {50},
    doi = {10.1214/21-AOS2152},
    issn = {0090-5364}
}

@article{Friedman2001GreedyMachine,
    title = {{Greedy Function Approximation: A Gradient Boosting Machine}},
    year = {2001},
    journal = {The Annals of Statistics},
    author = {Friedman, Jerome H},
    number = {5},
    pages = {1189--1232},
    volume = {29},
    publisher = {Institute of Mathematical Statistics},
    issn = {00905364}
}

@article{Angrist1996IdentificationVariables,
    title = {{Identification of Causal Effects Using Instrumental Variables}},
    year = {1996},
    journal = {Journal of the American Statistical Association},
    author = {Angrist, Joshua D. and Imbens, Guido W. and Rubin, Donald B.},
    number = {434},
    pages = {444--455},
    volume = {91},
    doi = {10.1080/01621459.1996.10476902},
    issn = {0162-1459}
}

@article{Bai2003InferentialDimensions,
    title = {{Inferential Theory for Factor Models of Large Dimensions}},
    year = {2003},
    journal = {Econometrica},
    author = {Bai, Jushan},
    number = {1},
    pages = {135--171},
    volume = {71},
    doi = {10.1111/1468-0262.00392},
    issn = {0012-9682}
}

@book{Bowden1990InstrumentalVariables,
    title = {{Instrumental Variables}},
    year = {1990},
    author = {Bowden, Roger John and Bowden, Roger J and Turkington, Darrell A},
    number = {},
    publisher = {Cambridge University Press},
    address = {Cambridge, United Kingdom}
}

@article{Cinelli2020MakingBias,
    title = {{Making Sense of Sensitivity: Extending Omitted Variable Bias}},
    year = {2020},
    journal = {Journal of the Royal Statistical Society Series B: Statistical Methodology},
    author = {Cinelli, Carlos and Hazlett, Chad},
    number = {1},
    pages = {39--67},
    volume = {82},
    doi = {10.1111/rssb.12348},
    issn = {1369-7412}
}

@article{Replogle2022MappingPerturb-SEQ,
    title = {{Mapping Information-Rich Genotype-Phenotype Landscapes with Genome-Scale Perturb-SEQ}},
    year = {2022},
    journal = {Cell},
    author = {Replogle, Joseph M and Saunders, Reuben A and Pogson, Angela N and Hussmann, Jeffrey A and Lenail, Alexander and Guna, Alina and Mascibroda, Lauren and Wagner, Eric J and Adelman, Karen and Lithwick-Yanai, Gila and Iremadze, Nika and Oberstrass, Florian and Lipson, Doron and Bonnar, Jessica L and Jost, Marco and Norman, Thomas M and Weissman, Jonathan S},
    number = {14},
    pages = {2559--2575},
    volume = {185},
    publisher = {Elsevier},
    doi = {10.1016/j.cell.2022.05.013},
    issn = {0092-8674}
}

@article{Wilms2021OmittedRelationships,
    title = {{Omitted Variable Bias: A Threat to Estimating Causal Relationships}},
    year = {2021},
    journal = {Methods in Psychology},
    author = {Wilms, R. and M{\"{a}}thner, E. and Winnen, L. and Lanwehr, R.},
    pages = {2590--2601},
    volume = {5},
    doi = {10.1016/j.metip.2021.100075},
    issn = {25902601}
}

@article{Hothorn2021PredictiveForests,
    title = {{Predictive Distribution Modeling Using Transformation Forests}},
    year = {2021},
    journal = {Journal of Computational and Graphical Statistics},
    author = {Hothorn, Torsten and Zeileis, Achim},
    number = {4},
    pages = {1181--1196},
    volume = {30},
    doi = {10.1080/10618600.2021.1872581},
    issn = {1061-8600}
}

@article{Meinshausen2006QuantileForests,
    title = {{Quantile Regression Forests}},
    year = {2006},
    journal = {Journal of Machine Learning Research},
    author = {Meinshausen, Nicolai},
    number = {35},
    pages = {983--999},
    volume = {7}
}

@article{Breiman2001RandomForests,
    title = {{Random Forests}},
    year = {2001},
    journal = {Machine Learning},
    author = {Breiman, Leo},
    number = {1},
    pages = {5--32},
    volume = {45},
    doi = {10.1023/A:1010933404324},
    issn = {08856125}
}

@article{Taylor2011RandomForests,
    title = {{Random Survival Forests}},
    year = {2011},
    journal = {Journal of Thoracic Oncology},
    author = {Taylor, Jeremy M.G.},
    number = {12},
    pages = {1974--1975},
    volume = {6},
    doi = {10.1097/JTO.0b013e318233d835},
    issn = {15560864}
}

@article{Wright2017Ranger:R,
    title = {{ranger: A Fast Implementation of Random Forests for High Dimensional Data in C++ and R}},
    year = {2017},
    journal = {Journal of Statistical Software},
    author = {Wright, Marvin N. and Ziegler, Andreas},
    number = {1},
    pages = {1--17},
    volume = {77},
    doi = {10.18637/jss.v077.i01},
    issn = {1548-7660}
}

@article{Stock2003Retrospectives:Regression,
    title = {{Retrospectives: Who Invented Instrumental Variable Regression?}},
    year = {2003},
    journal = {Journal of Economic Perspectives},
    author = {Stock, James H and Trebbi, Francesco},
    number = {3},
    pages = {177--194},
    volume = {17},
    publisher = {American Economic Association}
}

@misc{Ulmer2025SDModels:Models,
    title = {{SDModels: Spectrally Deconfounded Models}},
    year = {2025},
    booktitle = {10.32614/CRAN.package.SDModels},
    author = {Ulmer, Markus and Scheidegger, Cyrill},
    publisher = {10.32614/CRAN.package.SDModels},
    url = {https://cran.r-project.org/package=SDModels},
    doi = {10.32614/CRAN.package.SDModels}
}

@article{Scheidegger2025SpectralModels,
    title = {{Spectral Deconfounding for High-Dimensional Sparse Additive Models}},
    year = {2025},
    journal = {ACM / IMS Journal of Data Science},
    author = {Scheidegger, Cyrill and Guo, Zijian and B{\"{u}}hlmann, Peter},
    number = {1},
    volume = {2},
    publisher = {Association for Computing Machinery},
    address = {New York, NY, USA},
    doi = {10.1145/3711116}
}

@article{Cevid2020SpectralModels,
    title = {{Spectral Deconfounding via Perturbed Sparse Linear Models}},
    year = {2020},
    journal = {J. Mach. Learn. Res.},
    author = {{\'{C}}evid, Domagoj and B{\"{u}}hlmann, Peter and Meinshausen, Nicolai},
    number = {1},
    pages = {1--41},
    volume = {21},
    publisher = {JMLR.org},
    issn = {1532-4435}
}

@article{Meinshausen2010StabilitySelection,
    title = {{Stability Selection}},
    year = {2010},
    journal = {Journal of the Royal Statistical Society Series B: Statistical Methodology},
    author = {Meinshausen, Nicolai and B{\"{u}}hlmann, Peter},
    number = {4},
    pages = {417--473},
    volume = {72},
    doi = {10.1111/j.1467-9868.2010.00740.x},
    issn = {1369-7412}
}

@article{Hothorn2005SurvivalEnsembles,
    title = {{Survival Ensembles}},
    year = {2005},
    journal = {Biostatistics},
    author = {Hothorn, T.},
    number = {3},
    pages = {355--373},
    volume = {7},
    doi = {10.1093/biostatistics/kxj011},
    issn = {1465-4644}
}

@article{Gagnon-Bartsch2012UsingData,
    title = {{Using Control Genes to Correct for Unwanted Variation in Microarray Data}},
    year = {2012},
    journal = {Biostatistics},
    author = {Gagnon-Bartsch, J. A. and Speed, T. P.},
    number = {3},
    pages = {539--552},
    volume = {13},
    doi = {10.1093/biostatistics/kxr034},
    issn = {1465-4644}
}

\end{document}